\documentclass[journal,10pt]{IEEEtran}

\IEEEoverridecommandlockouts
\usepackage{cite}
\usepackage[justification=centering]{caption}
\usepackage{amsmath,amssymb,amsfonts}
\usepackage{algorithmic}
\usepackage{graphicx}
\usepackage{tabularx}
\usepackage{inputenc}
\usepackage{textcomp}
\usepackage{xcolor}
\usepackage{color}
\usepackage{stfloats}
\usepackage{multirow}
\usepackage{makecell}
\def\BibTeX{{\rm B\kern-.05em{\sc i\kern-.025em b}\kern-.08em
    T\kern-.1667em\lower.7ex\hbox{E}\kern-.125emX}}
\begin{document}

\title{Reconfigurable Intelligent Surfaces: Channel Characterization and Modeling}

\author{Jie Huang, \textit{Member, IEEE}, Cheng-Xiang Wang, \textit{Fellow, IEEE}, Yingzhuo Sun, Rui Feng, \textit{Member, IEEE}, Jialing~Huang, Bolun Guo, Zhimeng Zhong, and Tie Jun Cui, \textit{Fellow, IEEE} 
	\thanks{This work was supported by the National Key R\&D Program of China under Grant 2018YFB1801101, the National Natural Science Foundation of China (NSFC) under Grants 61960206006 and 61901109, the Frontiers Science Center for Mobile Information Communication and Security, the High Level Innovation and Entrepreneurial Research Team Program in Jiangsu, the High Level Innovation and Entrepreneurial Talent Introduction Program in Jiangsu, the High Level Innovation and Entrepreneurial Doctor Introduction Program in Jiangsu under Grant JSSCBS20210082, the Research Fund of National Mobile Communications Research Laboratory, Southeast University under Grant 2021B02, and the EU H2020 RISE TESTBED2 project under Grant~872172. \emph{(Corresponding author: Cheng-Xiang Wang.)}}
	\thanks{J. Huang, C.-X. Wang, Y. Sun, and  J.-L. Huang are with the National Mobile Communications Research Laboratory, School of Information Science and Engineering, Southeast University, Nanjing 210096, China, and also with the Purple Mountain Laboratories, Nanjing 211111, China (email: {\{j\_huang, chxwang, sunyingzhuo, jlhuang\}@seu.edu.cn}).}
	\thanks{R. Feng is with the Purple Mountain Laboratories, Nanjing 211111, China, and also with the School of Information Science and Engineering, Southeast University, Nanjing 210096, China (e-mail: fengxiurui604@163.com).}
	\thanks{B. Guo is with the Research Department of Beijing R\&D Subdivision, Wireless Network, Huawei Technologies Co., Ltd, Beijing 100085, China (email: guobolun@huawei.com).}
	\thanks{Z. Zhong is with the Wireless Network RAN Research Department, Huawei Technologies Co., Ltd, Shanghai 210206, China (email: zhongzhimeng@huawei.com).}
	\thanks{T. J. Cui is with the State Key Laboratory of Millimeter Wave, School of Information Science and Engineering, Southeast University, Nanjing 210096, China (email: tjcui@seu.edu.cn).}
}

\markboth{Proceedings of the IEEE,~Vol.~XX, No.~XX, MONTH~2021}%
{J. Huang \MakeLowercase{\textit{et al.}}: Bare Demo of IEEEtran.cls for IEEE Journals}

\maketitle

\begin{abstract}
Reconfigurable intelligent surfaces (RISs) are two dimensional (2D) metasurfaces which can intelligently manipulate electromagnetic waves by low-cost near passive reflecting elements. RIS is viewed as a potential key technology for the sixth generation (6G) wireless communication systems mainly due to its advantages in tuning wireless signals, thus smartly controlling propagation environments. In this paper, we aim at addressing channel characterization and modeling issues of RIS-assisted wireless communication systems. At first, the concept, principle, and potential applications of RIS are given. An overview of RIS based channel measurements and experiments is presented by classifying frequency bands, scenarios, system configurations, RIS constructions, experiment purposes, and channel observations. Then, RIS based channel characteristics are studied, including reflection and transmission, Doppler effect and multipath fading mitigation, channel reciprocity, channel hardening, rank improvement, far field and near field, etc. RIS based channel modeling works are investigated, including large-scale path loss models and small-scale multipath fading models. At last, future research directions related to RIS-assisted channels are also discussed. 
\end{abstract}


\section{Introduction}
Although the fifth generation (5G) wireless communication systems are being deployed worldwide, the performance metrics of 5G are not fully satisfied and the three application scenarios, i.e., enhanced mobile broadband (eMBB), massive machine type communications (mMTC), and ultra-reliable and low latency communication (URLLC) are not well supported. Hence, 5G will not meet all requirements of the future in 2030 and beyond \cite{You21}. Researches on the sixth generation (6G) wireless communication key technologies have been launched, with global coverage, all spectra, full applications, and strong network security as the vision and paradigm shifts \cite{You21, Wang20}. 

In the previous wireless communication systems, researchers are mainly focused on designing the transmitter (Tx), receiver (Rx), and corresponding algorithms to adapt to different propagation environments or wireless channels. Recently, reconfigurable intelligent surface (RIS), also known as intelligent reflecting surface (IRS), has been proposed and viewed as a potential key technology for the 6G wireless communication systems \cite{Bas19, Zhao19, Ren20_1, Gac20, Wu20, Gong20, Wu21, Yu21, Wong21, Yuan21, Pan21, Bas21, Zeng21, Liu21, Dai21, Long21, Bjo20_1, Liang21, Proc_Tat21}, since it can intelligently manipulate electromagnetic waves in a desired manner and smartly control propagation environments by low-cost and low-power near passive reflecting elements, thus improving the cost efficiency, energy efficiency, and spectral efficiency. 

In principle, RIS is a two dimensional (2D) metasurface consisting of massive elements (unit cells) with sub-wavelength distance, which is beyond the concept of massive multiple-input multiple-output (MIMO). Each element can be made by positive-intrinsic-negative (PIN) diodes, varactor diodes, liquid-crystals, \textcolor{black}{or graphene by implementing different fabrication methods. For example, the popular PIN/varactor diodes based RIS is composed of three layers. The outside layer has massive tunable metallic patches printed on a dielectric substrate to directly manipulate incident signals. The middle layer uses a copper panel to avoid the signal energy leakage. The inside layer is a control circuit board employed to tune the reflection coefficients of RIS elements in real time, which can be enabled via field programmable gate array (FPGA) \cite{Wu21, Pan21, Cui14_Light}. Other approaches include nano-network based method, where the RIS elements are connected to a communication nano-network and switched between different states to maximize the power sensed by the nano-network \cite{Lia19}.} 
 
\begin{figure}[t]
\begin{center}
\vspace{1.5cm}
 \includegraphics[width=3.5in]{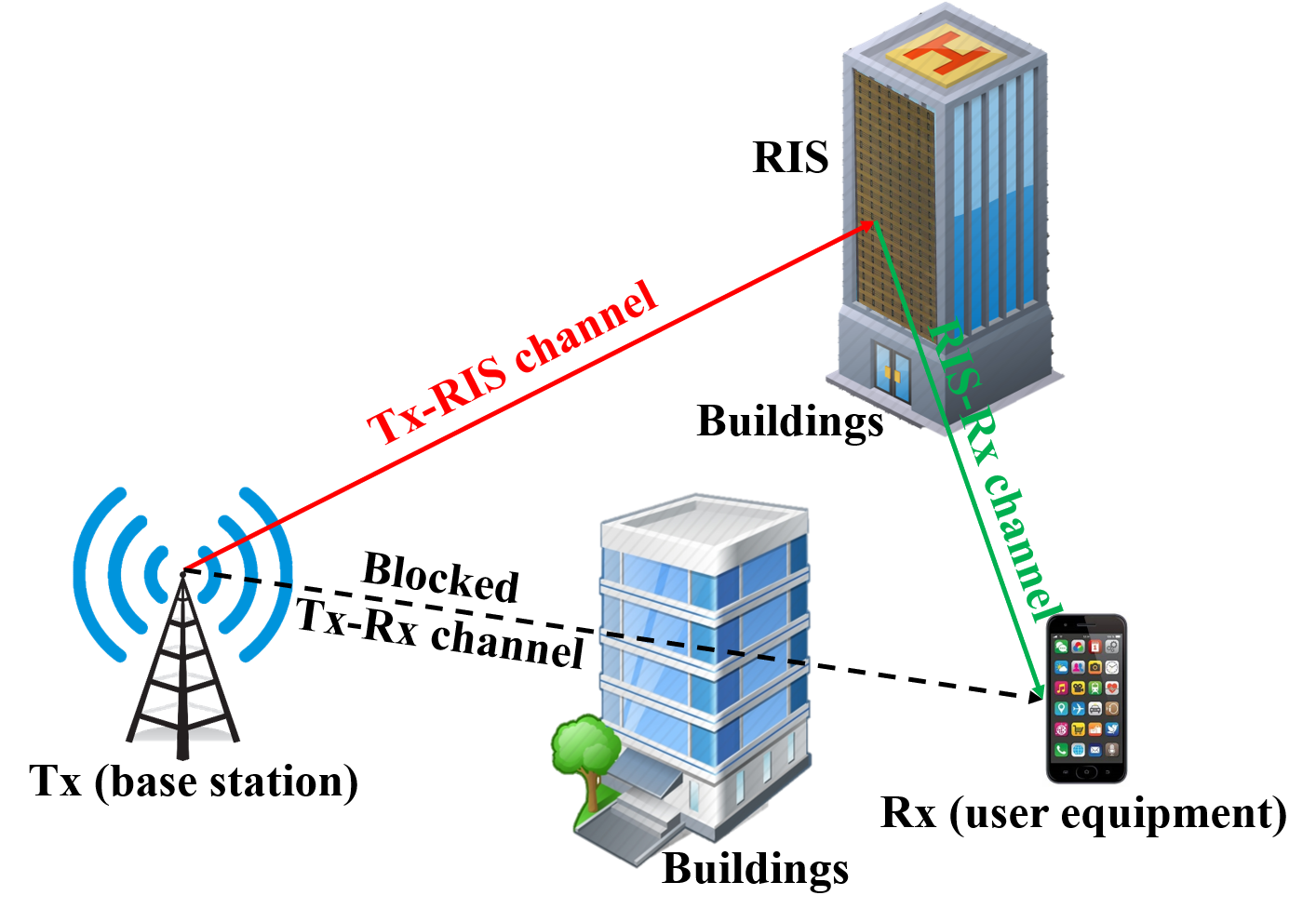}\\
  \caption{An illustration of the basic RIS-assisted communication scenario.}\label{RIS_scenario}
\end{center}
\end{figure}

While metasurface has been the frontier of physics in the last two
decades, its applications in wireless communications is just on the beginning \cite{Dai21}. Nevertheless, its advantages have attracted researchers in wireless communications to fully explore its potentials to meet the requirements of 6G. The basic communication scenario is that the line-of-sight (LOS) component between the Tx and the Rx is heavily blocked by objects such as buildings. In this case, an RIS can be employed to provide strong LOS links from Tx to RIS and from RIS to Rx, respectively, as shown in Fig. \ref{RIS_scenario}. A RIS can be easily deployed on internal walls of indoor environments, external facades of buildings, glasses of windows, etc. As the distance of the link via RIS is longer than the direct Tx-Rx link, a large number of RIS elements is essential to combat the higher path loss. 

\begin{figure*}[t]
\begin{center}
\vspace{1.5cm}
 \includegraphics[width=6in]{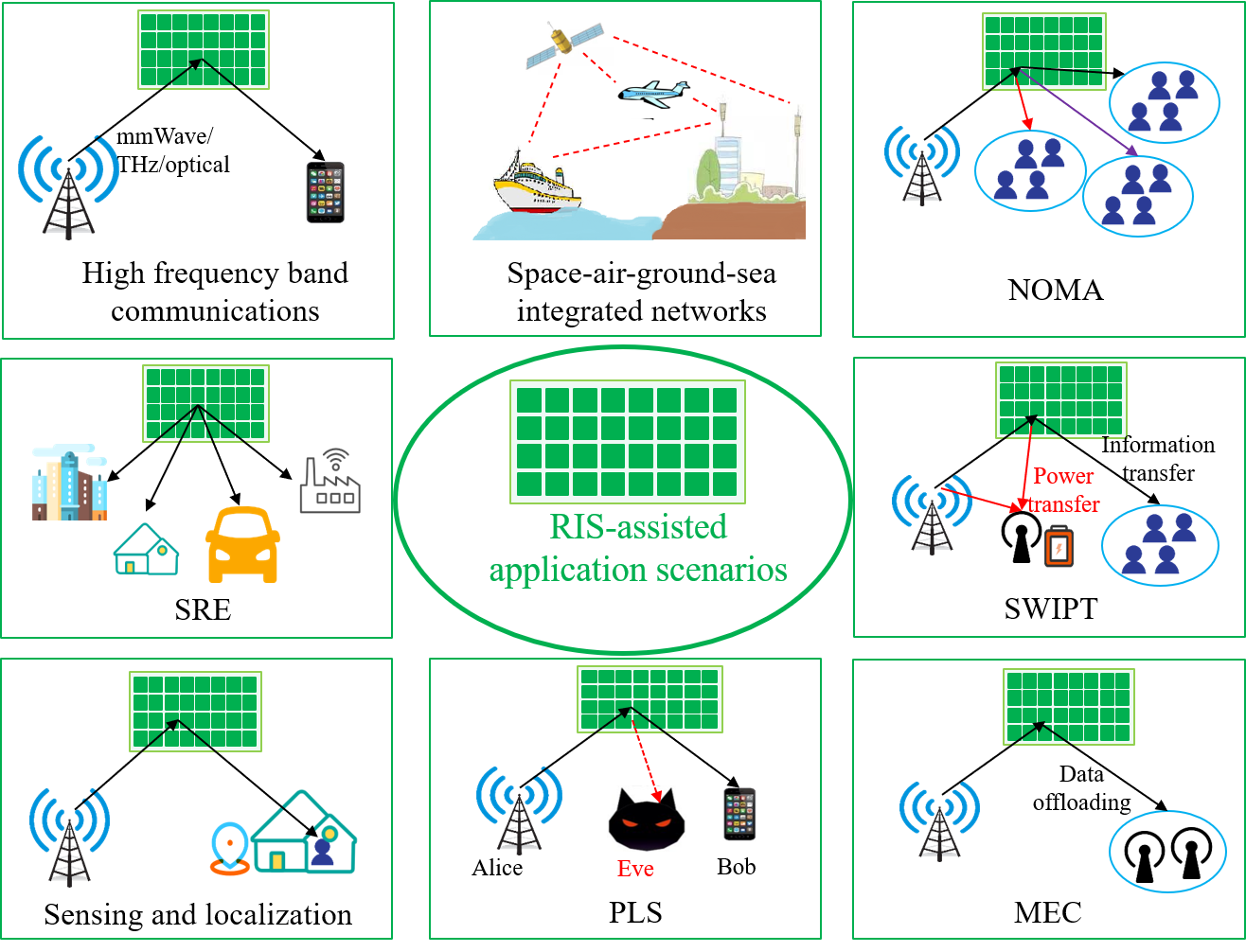}\\
  \caption{Potential RIS-assisted application scenarios.}\label{RIS_application}
\end{center}
\end{figure*}

Starting from the basic RIS-assisted communication scenario, many new applications are promising and under developments, as illustrated in Fig. \ref{RIS_application}. The potential application scenarios 
include:

1) High frequency band communications: All the spectra will be fully explored in the 6G era, especially the high frequency bands. As the frequency band increases, the wireless channel shows higher path loss, which limits the communication distance. \textcolor{black}{On one hand, the molecular absorption in high frequency bands aggravates the path loss. On the other hand, the high frequency signal will easily form an inaccessible shadow behind the obstacles due to the weak diffraction ability and serious penetration loss.} RIS can be well combined with high frequency band communications by creating alternative reflecting path and employing massive elements to enable high gains \textcolor{black}{and reconstruct propagation path towards the target user to avoid blockage}. Initial works on RIS-assisted  millimeter wave (mmWave), terahertz (THz) \cite{Chen21_1, Chen21_2, Wan21_2, Proc_Qu17, Proc_Sar21}, and optical \cite{Abu21} wireless communications are promising. \textcolor{black}{Since the optical wireless communication is mainly limited to LOS propagation and easily impaired by obstacles and geometric misalignment losses, RIS can help to confront non-LOS (NLOS) propagation by creating new links to improve the coverage range and reduce the outage probability. It showed that placing RIS closer to Rx can reduce outage probability when the jitter of Rx is more violent, while the middle of Tx and Rx is a better choice for other cases \cite{Naj21}. In addition, multi-branch RIS can further reduce outage probability compared with single-branch RIS \cite{Wangh21}.}

2) Space-air-ground-sea integrated networks: While 5G is mainly for terrestrial communications, 6G will be space-air-ground-sea integrated networks to enable the vision of global coverage. Satellite, unmanned aerial vehicle (UAV), and maritime communications will be integrated with terrestrial communications. The merits of RIS can be well exploited in the space-air-ground-sea integrated networks \textcolor{black}{to achieve a wide range of signal coverage}. As an example, for UAV communications, the trajectory mobility of UAV can create favorable propagation conditions, while the RIS can further improve the system performance \textcolor{black}{by being deployed on the ground to reflect signals from ground base stations to UAV \cite{You21_2}.} 

3) Non-orthogonal multiple access (NOMA): RIS-assisted NOMA communication systems can improve the spectral efficiency and user connectivity by opportunistically exploring the users’ different channel conditions \cite{Liu21}. Compared to conventional MIMO-NOMA systems, RIS-NOMA systems can overcome the challenges of the random fluctuation of wireless channels, blockage, and user mobility in an energy-efficient way \cite{Liu21}. 

4) Simultaneous wireless information and power transfer (SWIPT):
\textcolor{black}{Since RIS is a passive device, it can act as a relay-like function to reflect signals mixed with information and energy.} The drawback of low energy efficiency at the energy receivers can be solved with the aid of RIS. The wireless power transfer range can be enlarged and the number of required energy beams can be reduced, as shown in \cite{Liu21}.

5) Mobile edge computing (MEC): MEC is essential to offload the data traffic of wireless networks by employing the storage at edge servers to tackle the computation latency and guarantee the privacy and security of big data analysis~\cite{Yuan21}. \textcolor{black}{Because RIS can control the environment by adjusting the reflection coefficients, a beamforming gain can be achieved. Using this feature, in the edge network, RIS can increase the success rate of offloading of edge devices, thereby improving the performance of the entire network and reducing the end-to-end signal transmission delay. In addition, the adjustment efficiency of the RIS will further be improved with the utility of computing power of the edge server when RIS is deployed near it, thereby enhancing system coverage and transmission rate, as well as reducing edge network transmission and processing delays.}

6) Physical layer security (PLS): \textcolor{black}{As RIS can realize real-time reconfigurable electromagnetic environment and dynamic programmable wireless channel, it will further enhance the signal quality of intended users and decrease the interference signals from other users and then minimize the uncertainty of the  environment to the utmost extent. Therefore,} it is a potential key technology for PLS application\cite{Yuan21}.

7) Sensing and localization: Radio localization includes three parts, i.e., measurements, a reference system, and inference algorithms \cite{Wym20}. \textcolor{black}{RIS can be utilized to create a strong and consistent multipath in the propagation environment} and thus can improve the accuracy of localization.  \textcolor{black}{For example, RIS can support localization in harsh factory by circumventing LOS blockage.}

8) Smart radio environment (SRE): Due to its ability to intelligently control the propagation environments, RIS can be deployed to benefit our daily life, including smart city/home/transportation/factory \cite{Wu21}. In general, these applications are essential parts of SREs.

\textcolor{black}{We remark in each application scenario mentioned above, the specific RIS-assisted channel characteristics need to be considered in channel modeling. For example, in high frequency band communications, large bandwidth will bring frequency non-stationarity. In space-air-ground-sea integrated networks, temporal non-stationarity needs to be considered due to the movement of satellite and UAV, which will cause Doppler effect and multipath fast fading. When the size of RIS is large enough, the near field and spatial non-stationarity effects should also be considered.}
Current researches mainly concentrate on the theoretical analysis of performance improvements provided by RIS. However, the underlying wireless channel, which is crucial to the performance analysis, is not fully investigated. Many simplified assumptions in the theoretical analysis need to be validated by real RIS based channel measurements and experiments. Meanwhile, only a few of them take the RIS electromagnetic properties and their effects on the channel characteristics into account, such as the hardware impairments and mutual coupling between different elements. Moreover, most of the current works consider only the large-scale path loss, while the small-scale multipath fading is ignored or simplified to independent and identically distributed (i.i.d.) Rayleigh fading, which may lead to an overestimated performance gain. To fill these gaps, we focus on the detailed investigation of RIS based channel measurements and experiments, channel characteristics, and channel models, which are of importance to the design, optimization, and performance evaluation of RIS-assisted wireless communication systems. The novelty and main contributions of this paper are as follows.

\textcolor{black}{
\begin{itemize}
  \item [1)] 
  The available RIS based channel measurements and experiments are comprehensively surveyed, unveiling the new applications and performance gains of RIS-assisted wireless communications. Many new RIS based channel observations have been thoroughly analyzed.
  \item [2)]
  The various large-scale path loss models are investigated and compared, pointing out their applicabilities and limitations.
  \item [3)]
  The available small-scale multipath fading models are reviewed. The features of each modeling approach are discussed in details. 
\end{itemize}
}

The remainder of this paper is organized as follows. Section~II investigates the available RIS based channel measurements and experiments. In Section~III, the important channel characteristics brought by RIS are comprehensively surveyed. In Section~IV, the state-of-the-art channel modeling methods for RIS based channel are presented, including large-scale path loss models and small-scale multipath fading models. In Section~V, future research directions on RIS based channel measurements and modeling are discussed. Conclusions are drawn in Section~VI.

\section{RIS Channel Measurements and Experiments}

Recently, many types of RISs have been designed and fabricated for different purposes related to 6G research. Up to now, most of the RIS-assisted channel measurements and experiments in the literatures are conducted in relatively simple indoor environments at sub-6 GHz and mmWave bands. Although the prototypes and demonstrations are not enough, they are essential for the validation, theoretical analysis, and potential insights.

A key question is how to realize the RIS-assisted wireless communication systems. In \cite{Tang19}, a RIS-assisted wireless communication system prototype was presented. By changing the biasing voltage, the capacitance value of the varactor diode was adjusted and thus the reflection coefficient of each element was controlled, while the phase response was limited to 255$^\circ$. The system was based on national instruments (NI) universal software radio peripheral (USRP) platform. At the Tx side, the USRP generated a single tone incident electromagnetic wave to programmable RIS as the carrier signal. At the Rx side, the USRP down-mixed the received modulated radio frequency (RF) signal and sent the baseband signal to the host computer for synchronization and demodulation processing. Real-time video streaming experiment was conducted at 4 GHz in a typical indoor environment with RIS-Rx distance of 4 m, achieving 2.048 Mbps transmission rate. Experiment results showed that the QPSK constellation diagram after equalization was clear and stable, and the video stream can be transmitted smoothly and clearly even in the absence of channel coding. Then, the authors improved the design of RIS with a full 360$^\circ$ phase response \cite{Tang19_2}. Real-time video streaming experiment was conducted at 4.25 GHz in an indoor environment with RIS-Rx distance of 3 m, achieving 6.144 Mbps transmission rate by using 8PSK modulation. Furthermore, the real-time RIS-assisted 2$\times$2 MIMO-16QAM wireless communication experiment was conducted at 4.25 GHz in an indoor environment with RIS-Rx distance of 1.5 m \cite{Tang20_1, Tang20_2, Tang20_3, Tang20_4}. The 16QAM constellation diagrams of the two bitstreams recovered by the Rx were clear and dense, achieving 20 Mbps transmission rate. Since the reflection amplitude and phase responses of an element were coupled, a non-linear modulation technique was introduced to enable RIS-based QAM modulation under constant envelope constraint. In addition, the discrete phase shift was shown to be negligible when the quantization step was larger than 8.

As a comparison, two RISs working at 2.3 GHz and 28.5~GHz bands with 16$\times$16 elements were utilized for transmission demonstration on NI USRP platform in \cite{Dai20}. Measured in a compact anechoic chamber, a 21.7 dBi antenna gain was obtained by the RIS at 2.3 GHz, while a 19.1 dBi antenna gain was achieved at 28.5 GHz. A 20 m communication link was formed in an indoor environment, and a variety of modulation modes, including QPSK, 16QAM, 64QAM, etc, were employed, achieving 28.34 Mbps data transmission rate at 2.3 GHz.

In \cite{Zhang21}, a RIS-assisted wireless communication prototype was presented at 3.6 GHz. The RIS was capable of reflection and transmission. Each element was composed of two mirror symmetric layers. The NI USRP and directional double-ridged horn antennas were employed to transmit and receive signals at both sides. One Rx antenna was deployed at the same side, while the other Rx antenna was deployed on the opposite side in an indoor environment, where the walls were covered with wave absorbing materials to avoid the signals scattered from the environment. The results proved that the proposed beamforming scheme can generate directional beam towards the intended users on both sides of the RIS.

In \cite{Tang21_1}, path loss channel measurements were conducted in an anechoic chamber to validate the proposed path loss channel model. The Keysight E8267D signal generator, N9010A signal analyzer, two horn antennas, and three different RISs were utilized. The distances from Tx to RIS and from Rx to RIS were tuned to compare the measurement results with the proposed path loss model for the near field broadcasting and far field beamforming cases. In addition, new refined path loss models were proposed in \cite{Tang21_3} and corresponding mmWave channel measurements were conducted in an anechoic chamber. Specifically, two RISs were employed. The first one worked at 27 GHz, where the second one worked at 33 GHz. Two measurement systems were built. The first one was in an anechoic chamber to measure the impact of angle of reception on the power distribution of the signal reflected from the RIS. The second one was in a large empty room to measure the impacts of the distances of Tx-RIS and RIS-Rx. The new path loss models were validated. The authors also compared the received power of the 27 GHz RIS and a metal plate with the same shape and size in the anechoic chamber. The results showed that the RIS with a uniform coding state and a high reflection efficiency had the same reflection characteristics as a metal plate of the same shape and size.

In \cite{Tang21_2}, the aforementioned RISs at 4.25 GHz and 27 GHz were employed to study the channel reciprocity. The RIS at 4.25 GHz was made with varactor diodes and its phase can be continuously manipulated when the voltage of the RIS controller varied from 0~V to 21 V, whereas the RIS at 27 GHz was made with PIN diodes and its phase was only binary programmable. The uplink and downlink received complex signals were recorded in an indoor environment under different distances and angles from the Tx and Rx to the RIS, respectively. The RIS was also configured with identical, gradient, and stripe patterns. Good agreement between the uplink and downlink wireless channels was obtained, which validated the channel reciprocity of RIS-assisted time division duplex (TDD) wireless communication systems. 

In \cite{Pei21, Wang21_2}, a RIS at 5.8 GHz was employed for indoor and outdoor trials based on NI USRP platform. A simple channel reciprocity test was conducted by connecting two horn antennas to the two port R\&S vector network analyzer (VNA). Two horn antennas were placed close in front of the RIS to minimize the effects of multipath components (MPCs) other than the one reflected by the RIS. The measured $S_{12}$ and $S_{21}$ parameters represented the uplink and downlink channels, respectively. Similar to \cite{Tang21_2}, the amplitudes and phases of the two channels showed consistent, which validated the channel reciprocity. In addition, initial indoor and outdoor \textcolor{black}{over the air (OTA)} tests were carried out. In the indoor test, about 26 dBi gain was achieved by using RIS. In outdoor tests, about 27 dBi and 14 dBi gains were obtained for 50 m and 500 m links, respectively. The larger distance had performance degradation as the wireless channel became more complicated.

In \cite{Mud21}, RIS-assisted mmWave channel measurements were conducted in indoor laboratory and office environments. Different from the conventional case where the RIS is between the Tx and Rx, the RIS was utilized as Tx. It was based on linearly polarized binary-phase unit cells with 180$^\circ$ phase-shifting capability. The phase shift provided by each unit cell through a proper bias of the PIN diodes was computed using an optimization tool. The channel sounder consists of a four-ports VNA, the RIS, one monopole antenna, and an antenna positioner. Channel measurements were conducted at 28 GHz with 4 GHz bandwidth. The results showed that antenna beamforming can largely reduce the path loss and delay spread in the main beam directions, which led to an improvement of the link budget. 

In \cite{Far21_1, Far21_2}, experiments were conducted to assess the RIS-assisted ambient backscatter communications. The source, the tag, and the reader were all equipped with dipole antennas. The USRP was used at both sides. The transmitted signal was at 5.4 GHz with 500 MHz bandwidth. The reader received the combined signals of the source, the RIS, and the tag. As the RIS controlled the reflected signal, a hot spot or coherent spots on the tag and the reader can be created to assist the communication between the tag and the reader.

In \cite{Zhang21_1}, the concept of active RIS was proposed to break the double fading effect of passive RIS. Different from the existing passive RIS that reflects signals passively without amplification, active RIS can actively amplify the reflected signals. Experiments were conducted by using an active RIS at 2.36 GHz. The incident signal and the pump input were coupled in a varactor-diode-based reflection-type amplifier to generate the reflected signal with amplification.



In \cite{Gro21}, a RIS was fabricated at 28 GHz based on a binary-phase tunable metasurface. Experiments were conducted in near field and far field cases. In near field case, the RIS was designed to be a typical reflectarray antenna and proved to be able to beamform mmWave signals very efficiently, achieving directivity around 30 dBi and reaching angles as high as 60$^\circ$ as predicted by the analytical model. In far field case, two horn antennas were connected to a VNA without LOS propagation. The RIS was proved to be able to create a LOS propagation, which provided a 25 dBi gain over the received power.

Based on the design described in \cite{Gro21}, a RIS was fabricated at 28 GHz in \cite{Pop21}. Four RISs with unit cells of 20$\times$20 were assembled to achieve 40$\times$40 unit cells. Indoor measurements were conducted to demonstrate the functions of RIS. The RIS was placed on a wall and two horn antennas were used at the Tx and Rx sides to emulate the base station and user equipment, respectively. The Tx-RIS and RIS-Rx distances were 5.5 m and 2 m, respectively. To create a RIS-assisted LOS path, the positions of Tx and Rx were estimated in the beamforming algorithm. A 30 dBi gain was achieved in a wide frequency range of 3 GHz. In addition, NI USRP platform was utilized for QPSK signal transmission, where the four modulated symbols were well distinguished.

In \cite{Wan21}, a prototype of the binary-phase metasurface was designed and tested. The simulation showed that the reflection amplitudes had little changes but the reflection phases were opposite at 28 GHz when the PIN diode switches. The functions of steerable sum and difference beams were realized. In addition, the beemsteering performance was analyzed in \cite{Li20, Li21_1, Li21_2, Li21_3}, which can achieve precise tracking with an angle mismatch of 0.5$^\circ$-2.5$^\circ$ and minimizes the beam training overhead through flexible beam refinement.

In \cite{Ale21}, experiments were conducted for RIS-assisted wireless communications in rich scattering environments. Two cases were studied. First, an RIS was deployed to shape the channel impulse response (CIR) to enable higher achievable communication rates. Second, the RIS-tunable propagation environment was leveraged as an analog multiplexer to localize non-cooperative objects using wave fingerprints, even when they were outside the LOS propagation region.

In \cite{Sta21}, measurements were conducted to show that RIS can enable attackers to perform complex signal manipulation attacks on the physical layer in real time with low-cost. A laptop was connected to the network to measure the end-to-end connection quality. The proposed environment reconfiguration attack scheme was able to largely reduce the throughput.

In \cite{Chu21}, a RIS at 28 GHz was designed to perform as reflectarray and transmitarray on each side. Five functions were achieved, including reflection magnitude, transmission magnitude, reflection phase difference, transmission phase difference, and polarization conversion ratio. The channel reciprocity was also validated by connecting two horn antennas to a VNA and put the RIS between them.

A summary of the above-mentioned RIS channel measurements and experiments is given in Table \ref{tab:prototype}, \textcolor{black}{classifying the frequency bands, scenarios, system configurations, RIS constructions, experiment results, and channel observations}. As can be seen, initial channel measurements and experiments have unveiled various characteristics of RIS-assisted wireless communication systems, such as the transmission rate, received signal power, path loss, and channel reciprocity. Meanwhile, many new applications are proved to be possible, including RIS-assisted ambient backscatter communications, active RIS, RIS as a transmitarray/reflectarray, RIS for physical layer attack, RIS-assisted rich scattering environment, etc.

\begin{table*}[tb!]
\caption{\textcolor{black}{A summary of RIS experiments, channel measurements, and channel observations at sub-6 GHz and mmWave bands.}}
\label{tab:prototype}
\centering
  \arraybackslash
\begin{tabular}{|m{0.6cm}<{\centering}|m{1.3cm}<{\centering}|m{1.4cm}|m{2.3cm}|m{3.4cm}|m{4.3cm}|m{2cm}|}
\hline
\textbf{Ref.}&\textbf{Frequency band (GHz)}&\textbf{Scenario}&\textbf{System configuration}&\textbf{RIS information}&\textbf{Experiment results}&\textbf{Channel observation}\\
\hline
\cite{Tang19}& 4 & Indoor&NI USRP platform&\thead{Elements: 16$\times$8;\\ Size: 252.8$\times$176~mm$^2$} & QPSK transmission rate of 2.048 Mbps with RIS-Rx distance of 4 m&Transmission rate\\
\hline
\cite{Tang19_2}& 4.25 & Indoor&NI USRP platform&\thead{Elements: 32$\times$8;\\ Size: 5.44$\lambda\times$1.36$\lambda$ \\($\lambda$: wavelength)}& 8PSK transmission rate of 6.144 Mbps with RIS-Rx distance of 3 m&Transmission rate\\
\hline
\cite{Tang20_1, Tang20_2, Tang20_3, Tang20_4}& 4.25 & Indoor& NI USRP platform&\thead{Elements: 32$\times$8;\\ Size: 5.44$\lambda\times$1.36$\lambda$}  & 2$\times$2 MIMO-16QAM transmission rate of 20 Mbps with RIS-Rx distance of 1.5 m&Transmission rate\\
\hline
\cite{Dai20}& 2.3, 28.5 & Indoor&NI USRP platform&\thead{Elements: 16$\times$16; \\Size: 6.13$\lambda\times$6.13$\lambda$ (2.3 GHz)}&  QPSK, 16QAM, and 64QAM transmission, achieving 28.34 Mbps data transmission rate at 2.3 GHz&Transmission rate\\
\hline
\cite{Zhang21}& 3.6 & Indoor&NI USRP platform &\thead{Elements: 640; \\Size: 2.87$\times$1.42$\times$0.71 cm$^3$\\ for each element}& Demonstrate the ability to beamform towards the intended users on both sides of the RIS&-\\
\hline
\cite{Tang21_1}& 4.25, 10.5 &Anechoic chamber&Keysight signal generator and signal analyzer&Elements (4.25/10.5/ 10.5~GHz): 32$\times$8/ 100$\times$102/ 50$\times$34 ; Size (4.25/10.5/10.5 GHz): 5.44$\lambda\times$1.36$\lambda$/ 35$\lambda\times$35.7$\lambda$/ 17.5$\lambda\times$11.9$\lambda$&1) Develop analytical free-space path loss models for RIS-assisted wireless communications; 2) Conduct path loss channel measurements and results match well with modeling&Received signal power\\
\hline
\cite{Tang21_3}& 27, 33 &Anechoic chamber and indoor&Keysight signal generator and signal analyzer&Elements (27/33 GHz): 56$\times$20/ 40$\times$40; Size (27/33 GHz): 78.4$\times$56.0 mm$^2$/ 152$\times$152 mm$^2$ &Refine path loss model and validate its accuracy in the sub-6 GHz and mmWave frequency bands through path loss channel measurements&Received signal power\\
\hline
\cite{Tang21_2}& 4.25, 27 &Indoor&Keysight signal generator and signal analyzer &Elements (4.25/27 GHz): 32$\times$8/ 56$\times$20; Size (4.25/27~GHz): 5.44$\lambda\times$1.36$\lambda$/ 7.06$\lambda\times$5.04$\lambda$&Validate the channel reciprocity of RIS assisted TDD wireless communication systems&Channel reciprocity \\
\hline
\cite{Pei21, Wang21_2}& 5.8 & Indoor and outdoor&R\&S VNA, NI USRP platform & \thead{Elements: 55$\times$20;\\ Size: 15.48$\lambda\times$6.05$\lambda$}&Validate channel reciprocity and demonstrate RIS-assisted transmission &Channel reciprocity\\
\hline
\cite{Mud21}&28 &Indoor laboratory and office &R\&S VNA&\thead{Elements: 20$\times$20; \\Size: 9.52$\lambda\times$9.52$\lambda$}&Demonstrate Tx employing RIS to improve the link budget &Path loss and delay spread\\
\hline
\cite{Far21_1, Far21_2}& 5.4 & Indoor&NI USRP platform &\thead{Elements: 14$\times$14; \\Size: 14$\times$14 mm$^2$ \\for each element}&  Assess the RIS-assisted ambient backscatter communications&Bit error rate (BER)\\
\hline
\cite{Zhang21_1}& 2.36 & Indoor&VNA and spectrum analyzer&\thead{Active RIS}&  Demonstrate the active RIS to break the fundamental limit of the double fading effect&Channel capacity\\
\hline
\cite{Gro21}& 28 & Indoor&VNA & \thead{Elements: 20$\times$20;\\ Size: 10$\times$10 cm$^2$}& Use RIS to beamform mmWave signals as a reflectarray antenna in near field and to create a LOS propagation for signal improvement in far field&Received signal power\\
\hline
\cite{Pop21}& 28 & Indoor&VNA, NI USRP platform &\thead{Elements: 40$\times$40;\\ Size: 20$\times$20 cm$^2$}&  Prove RIS's ability to create a LOS propagation for signal improvement and transmit QPSK modulated signals&Received signal power\\
\hline
\cite{Wan21, Li20, Li21_1, Li21_2, Li21_3}& 28 &Indoor&NI USRP platform &  \thead{Elements: 20$\times$20}& 1) Prove the RIS's functions to steerable sum and difference beams; 2) Prove the RIS's function of beemsteering&-\\
\hline
\cite{Ale21}& 2.5 & Indoor&-&\thead{Elements: 47}& Prove the RIS's functions to shape the CIR and localize non-cooperative objects using wave fingerprints&Transmission rate\\
\hline
\cite{Sta21}& 5.35 & Indoor&WLAN router &\thead{Elements: 16$\times$8;\\ Size: 40$\times$16 cm$^2$}&  Demonstrate that RIS can enable attackers to perform complex signal manipulation attacks on the physical layer in real time with low-cost&-\\
\hline
\cite{Chu21}& 28 &Indoor& VNA & \thead{Elements: 15$\times$15;\\ Size: 60$\times$60 mm$^2$}& 1) Prove the functions of reflection magnitude, transmission magnitude, reflection phase difference, transmission phase difference, and polarization conversion ratio; 2) Validate the channel reciprocity&Channel reciprocity\\
\hline
\end{tabular}
\end{table*}

\section{RIS Channel Characteristics}
As RIS has the ability to tune the amplitude and/or the phase of incident signals, many new channel characteristics will appear, which are substantially different from conventional wireless channels. \textcolor{black}{Some of the specific channel characteristics have been validated by channel measurements and experiments, while others are theoretically analyzed.} Here, we mainly introduce channel characteristics of reflection and transmission, Doppler effect and multipath fading mitigation, channel reciprocity, channel hardening, rank improvement, near field and far field, and half-power beamwidth (HPBW) and aperture efficiency. Other new channel characteristics include: multipath labeling \cite{Wang21_3}, which intended to inject a label on propagation paths through the RISs; OTA equalization \cite{Ars21}, which adjusted the phases of RIS elements according to the incoming signals to maximize the magnitude of the first tap and made the inter-symbol interference negligible; electromagnetic interference \cite{Tor21}, which had a non-negligible impact on communication performance, especially when the size of the RIS grew large; etc.

\subsection{Reflection and transmission}
In traditional wireless communication systems, transmission signals will have a large penetration loss caused by objects, which can be up to \textcolor{black}{several tens of dB}, especially for mmWave and THz bands. Thus, the wireless signals are mainly received by the Rx antenna via reflection, scattering, and diffraction. 

For RIS-assisted communications, the RIS can be configured in the reflecting mode, the transmitting mode, or simultaneously transmitting and reflecting (STAR) mode \cite{Zeng21, Xu21, Liu21_2, Mu21}. In the reflecting mode, the users at the transmitter's same half-space are served, while the users at the other half-space are served in the transmission mode. In the simultaneously reflection and transmission mode, the users in the full space are able to be served. Preliminary simulation results showed that STAR-RISs can significantly reduce the base station power consumption compared to conventional reflecting/transmitting-only RISs \cite{Liu21_2, Mu21}. Up to now, most of the works on RIS-assisted communications mainly focus on the reflecting mode, while the other two modes need further investigation.
 
\subsection{Doppler effect and multipath fading mitigation}
When the user equipment is moving, it will cause Doppler effect and fast multipath fading. As the RIS can tune the phase of incident signals, it can align the phases of MPCs arriving from different directions, i.e., the rapid fluctuations in the received signal strength due to the Doppler effect can be effectively reduced by using the real-time tunable RIS \cite{Bas21_1}. The simulation results proved that the multipath fading effect can be totally eliminated when all reflectors in a propagation environment are coated with RIS, while even a few RIS can significantly reduce the Doppler spread as well as the deep fading in the received signal for general propagation environments with several interacting objects \cite{Bas21_1}. In \cite{Mat21}, the authors showed that the received power can be maximized without adding Doppler spread to the system.

In \cite{Wang21}, a RIS-assisted high-mobility vehicular communication system was considered. The RIS is fixed between the roadside BS and high-mobility vehicle. The predictable information of vehicle, such as speed, trajectory, and timetable was utilized to design a sub-optimal phase shift set of RIS, which made the RIS capable of maximizing instantaneous signal-to-noise ratio (SNR), minimizing Doppler spread, and keeping low delay spread simultaneously.

Different from \cite{Wang21}, a RIS-assisted high-mobility vehicular communication system was studied in \cite{Huang21_1, Huang21_2}, where the RIS was mounted on the top of high-mobility vehicle. An efficient two-stage transmission protocol was proposed to achieve RIS channel estimation and refraction optimization for data transmission by exploiting the quasi-static channel between the high-mobility RIS and user as well as the LoS dominant channel between the BS and RIS. Simulation results showed that the proposed protocol can achieve the full RIS passive beamforming gain and convert the overall BS-user channel from fast to slow fading to enable reliable transmission. 

\subsection{Channel reciprocity}
Channel reciprocity means that the wireless channel from the Tx side to the Rx side is the same with the wireless channel from the Rx side to the Tx side \cite{Proc_Asa20}. This phenomenon comes from the symmetry of Maxwell's equations with respect to time~\cite{Tang21_2}. In general, when the control signals applied to the unit cells remain unchanged, commonly designed and fabricated RISs inherently obey the reciprocity theorem. This is because the materials that constitute RISs, such as metal patches, dielectric layers, and electronic components, often conform to the Rayleigh-Carson reciprocity theorem \cite{Tang21_1, Tang21_2}. The experiments at 4.25 GHz and 27 GHz bands demonstrate channel reciprocity, which show consistent between uplink and downlink channels. In addition, by using active nonreciprocal circuits, performing time-varying controls, and employing nonlinearities and structural asymmetries, the RIS channel reciprocity can be broken. An example of reciprocity breaking by RIS was shown in \cite{Dai21}. The channel reciprocity was also validated in \cite{Pei21} and \cite{Chu21}.

An angle-dependent phase shifter model for varactor-based RISs was proposed based on the equivalent circuit in \cite{Chen20}. The simulation results indicated that the angle-reciprocity of varactor-based RIS only held under small incident angles of both forward and reverse incident electromagnetic waves, thus limited the channel reciprocity in RIS-assisted TDD systems.

In \cite{Luo21}, a RIS-assisted pilot decontamination scheme was proposed for TDD massive MIMO system by breaking the channel reciprocity. Specifically, the phases of RIS elements were configured differently during the uplink transmission and downlink transmission. In this case, the inter-cell interference due to the RIS-assisted link can be mitigated while the non-RIS link interference still existed. If there was only the RIS-assisted link between the  Tx and Rx, by increasing the RIS elements and randomly configuring the phases, the pilot contamination can be mitigated asymptotically. The system performance was then proved to be significantly improved.

\subsection{Channel hardening}
MIMO channels can provide spatial diversity to reduce SNR variations. In particular, i.i.d. Rayleigh fading channels give rise to channel hardening, where the SNR variations average out as the number of antennas increase. In \cite{Bjo20_2}, the authors provided a new definition of channel hardening in RIS-assisted communications, which stated that the ratio between the system SNR and the square of the number of RIS elements tends to be a constant value when the number of RIS elements tends to infinity. The simulation results revealed that the random SNR variations in the case of random phases remain large since there is no hardening. When the direct path is present, the SNR increases very slowly with the number of RIS elements. Hence, an RIS should be properly configured to benefit from the channel hardening and SNR gain. 

In \cite{Chi21}, the channel hardening effect was investigated for RIS-assisted massive MIMO system. Once the channel hardening effect is achieved, the effective channel gain can be replaced by its mean value. The base station can replace the instantaneous channel gain by its mean value for signal detection in the uplink. In the downlink, each user can treat the mean value of the effective channel gain as the true one to detect the desired signal and no downlink pilot overhead is required for channel estimation.

In \cite{Jun20}, the channel hardening effect was theoretically analyzed and the performance bound was asymptotically derived for the RIS-assisted system. The derived ergodic rates showed that
hardware impairments, noise, and interference from estimation errors and the NLOS path became negligible as the number of antennas increases.

\subsection{Rank improvement}
In point-to-point MIMO communications, large spatial multiplexing gains can be achieved at the high SNR region, while high SNR mainly appears in LOS scenarios where the channel matrix has low rank and therefore does not support spatial multiplexing \cite{Ozd20_1}. In this case, RIS can enrich the propagation environment by adding MPCs with distinctively different spatial angles, thus a multiplexing gain is achieved even when the direct path has low rank. The simulation results in \cite{Ozd20_1} revealed that the performance greatly depends on the path loss and deployment angles. 

The rank deficiency can be analyzed in the form of the eigenvalues of the covariance matrix \cite{Chi21}. The RIS was proved to be able to compensate for the rank deficiency appearing in massive MIMO communications.

To solve the problem of rank deficiency due to the unobstructed LOS in UAV communications, RIS was utilized to enrich the channel in \cite{Ozt21}. An RIS placement approach was proposed to improve the spatial multiplexing gains and maximize the average channel capacity in a predefined drone corridor. The RIS-assisted channel was shown to have similar performance at sub-6 GHz and mmWave bands.

\subsection{Near field and far field}
In the far field, the total channel gain grows linearly with the number of antenna elements $N$, while it grows as $N^2$ for RIS \cite{Bjo19, Bjo20_1, Bjo20_3}. However, the growth rate eventually tapers off when entering the near field. \textcolor{black}{In general, the Rayleigh distance criterion is applied to distinguish the near field and far field cases. In near field, the spherical wavefront effect needs to be modeled, while the plane wavefront can be assumed in far field. Meanwhile, the distances and geometry relationship for each pair of Tx/Rx and RIS elements will influence the electromagnetic response of RIS-assisted channel directly.}
An exact near field channel model is needed to evaluate the system performance whenever the distance between the Tx/Rx and RIS is comparable with the dimension of RIS \cite{Tor20}.

\subsection{HPBW and aperture efficiency}
The HPBW of RIS was investigated in uniform linear array (ULA) and uniform rectangular array (URA) configurations under the far field condition without considering path loss and fading\cite{Han21}. It showed that RIS's HPBW is equal to antenna array's HPBW at the same reflect/transmit angle with maximal ratio combining applied, while it is greater than or equal to antenna array's HPBW when all weights are equal. Another feature is that RIS's HPBW will change with the incident angles. In addition, the aperture efficiency of RIS was studied in \cite{Cho21}.

\section{RIS Channel Modeling}

\subsection{Large-scale path loss models}
As the RIS-assisted channel is the cascade of three sub-channels, the large-scale path loss models are affected by the electromagnetic properties of RIS and show great differences with traditional channels. Several path loss models have been proposed for different scenarios and taken different parameters into account. A comprehensive survey of the path loss models are given below.

In \cite{Bas19}, the authors illustrated the mechanism of RIS by revisiting the two-ray path loss model, which consisted of a LOS ray and a ground-reflected ray. The RIS with $Q$ elements was assumed to be laid on the ground to assist the communications. The received power was approximated as

\begin{equation}
P_r\approx (Q+1)^2P_t(\frac{\lambda }{4\pi d_{tr}})^2
\end{equation}
where $P_t$ is the transmitted power, and $d_{tr}$ is the Tx-Rx distance. The received power is proportional to $Q^2$ and decay with the inverse of $(d_{tr})^2$. By increasing the RIS elements, the path loss is able to be combated.

In \cite{Ozd20_2}, the authors proposed a far field path loss model for RIS communications based on physical optics technique. It also explained why the surface consists of many elements that individually act as diffuse scatterers but can jointly beamform the signal in a desired direction with a certain beamwidth. Specifically, the far field path loss was expressed as

\textcolor{black}{
\begin{equation}
PL=\frac{(4\pi d_1d_2)^2}{G_tG_r(ab)^2\textup{cos}^2{\theta _i}\textup{sinc}(\frac{\pi b}{\lambda }(\textup{sin}\theta _s-\textup{sin}\theta _r))}
\end{equation}
where $\textup{sinc}(x) = \textup{sin}(x)/x$ is the sampling function}, $a$ and $b$ are the sizes of the rectangular RIS. $G_t$ and $G_r$ are the Tx and Rx antenna gains, respectively. $d_1$ and $d_2$ are the distances from the Tx and Rx to the RIS, respectively. $\theta_i$, $\theta_r$, and $\theta_s$ are the incident angle, desired reflection angle, and observed reflection angle, respectively. When we observe at the desired reflection angle, i.e., $\theta_s$ = $\theta_r$, the path loss model simplifies to

\begin{equation}
PL=\frac{(4\pi d_1d_2)^2}{G_tG_r(ab)^2\textup{cos}^2{\theta _i}}.
\end{equation}

The received power is proportional to the square of the RIS area and $1/(d_1d_2)^2$, which \textcolor{black}{was different from} the conjecture in \cite{Bas19} that the received power would be proportional to $1/(d_1 + d_2)^2$. That conjecture might hold for an infinitely large RIS or in the near field, if the RIS is configured to act as a mirror, but provably not in the far field setup.

In \cite{Bjo19}, the authors compared conventional massive MIMO with RIS-assisted communications and proved that the RIS-assisted communications can never provide higher information rate or SNR. Numerical results showed that thousands of reflecting elements are needed as the RIS can not amplify the transmitted signal. 

A general RIS-assisted path loss model was proposed in \cite{Tang21_1}, which was expressed as

\begin{equation}
\begin{split}
PL=&\frac{64\pi ^{3} }{G_tG_rGd_xd_y\lambda^2 }\times\\
&\left | \sum\limits_{m=1}^{M}\sum\limits_{n=1}^{N}\frac{\sqrt{F_{n,m}^{combine}}\Gamma_{n,m}}{r_{n,m}^{t}r_{n,m}^{r}} e^{\frac{-j2\pi (r_{n,m}^{t}+r_{n,m}^{r})}{\lambda }}\right |^{-2}
\end{split}
\end{equation}
where $d_x$ and $d_y$ are the size of each element along x axis and y axis, respectively. $M$ and $N$ are the number of elements along x axis and y axis, respectively. $G$ is the gain of one element, $F_{n,m}^{combine}$ denotes the effect of the normalized power radiation pattern of each element on the received signal power, $\Gamma_{n,m}$ is the reflection coefficient of each element with amplitude of $A$ and phase of $\phi$, and $r_{n,m}^{t}$ and $r_{n,m}^{r}$ denote the distances from the Tx and Rx to each element. As we can see, the received power is related to the Tx/Rx antenna gains, the gain, size, number, and specific electromagnetic properties of elements, and the travel distances. In addition, the general path loss model was simplified to three cases, i.e., far field beamforming (both Tx and Rx are in the far field), near field beamforming (at least one in the near field), and near field broadcasting (at least one in the near field). 

In far field beamforming case, the path loss was expressed as 

\begin{equation}
PL=\frac{64\pi ^{3}(d_1d_2)^2 }{G_tG_rGM^2N^2d_xd_y\lambda^2 F(\theta_t,\phi_t)F(\theta_r,\phi_r)A^2}
\end{equation}
where $F(\theta_t,\phi_t)$ and $F(\theta_r,\phi_r)$ are the normalized radiation patterns of each element at the incident direction ($\theta_t$,$\phi_t$) and the reflection direction ($\theta_r$,$\phi_r$). 

In near field beamforming case, the distances of Tx and Rx to each element is different and the corresponding path loss was expressed as 

\begin{equation}
PL=\frac{64\pi ^{3} }{G_tG_rGd_xd_y\lambda^2A^2\left | \sum\limits_{m=1}^{M}\sum\limits_{n=1}^{N}\frac{\sqrt{F_{n,m}^{combine}}}{r_{n,m}^{t}r_{n,m}^{r}} \right |^2 }.
\end{equation}

In the near field broadcasting case, the path loss was approximated as 

\begin{equation}
PL\approx \frac{16\pi ^{2}(d_1+d_2)^2 }{G_tG_r\lambda^2A^2 }.
\end{equation}

Note that for each case, the phase shift of each element was differently designed. The proposed path loss models were also validated by channel measurements at 4.25 GHz and 10.5~GHz in an anechoic chamber.
 
In \cite{Tang21_3}, the authors further refined their proposed path loss models in \cite{Tang21_1} to make it easier to use. The impact of the radiation patterns of the antennas and RIS elements was considered as an angle-dependent loss factor. In the far field beamforming case, the path loss model was refined as 

\begin{equation}
PL=\frac{16\pi ^{2}(d_1d_2)^2 }{G_tG_rG(MNd_xd_y)^2 F(\theta_t,\phi_t)F(\theta_r,\phi_r)A^2}.
\end{equation}

The near field beamforming path loss was refined as

\begin{equation}
PL=\frac{16\pi ^{2} }{G_tG_r(d_xd_y)^2\left | \sum\limits_{m=1}^{M}\sum\limits_{n=1}^{N}\frac{\sqrt{F_{n,m}^{combine}}}{r_{n,m}^{t}r_{n,m}^{r}} \right |^2}.
\end{equation}

In addition, the path loss model for one element was proposed as

\begin{equation}
PL=\frac{16\pi ^{2} (r_{n,m}^{t}r_{n,m}^{r})^2}{G_tG_r(d_xd_y)^2F_{n,m}^{combine}\left | \Gamma_{n,m} \right |^2}.
\end{equation}

The refined path loss models were then validated by channel measurements at 27 GHz and 33 GHz bands in both anechoic chamber and indoor environments.

In \cite{Ell19}, a path loss model for RIS communications was proposed by considering the number of elements, the physical dimensions of the RIS, and the radiation pattern and spacing of constituent elements. It was expressed as

\begin{equation}
PL=\frac{256\pi ^{4}}{\eta G_tG_r}\left | \sum\limits_{q=1}^{Q}A_n\frac{\sqrt{G(\theta_t,\phi_t)G(\theta_r,\phi_r)}}{d_{1n}d_{2n}} e^{j\phi_n}\right |^{-2}
\end{equation}
where $\eta$ is the power efficiency of RIS, which is less than 1 for 
passive RIS. $A_n$ and $\phi_n$ are the amplitude and phase of the $n$th element, respectively. $d_{1n}$ and $d_{2n}$ are the distances from the Tx and Rx to the $n$th element, respectively. The path loss expression for far field scenario was also derived. The proposed path loss model was then applied to a reflectarray-type RIS, yielding insights into performance as a function of the size of the RIS, proximity of the RIS to the Tx and Rx, and the criteria used to control the elements. 

In \cite{Gar20}, the authors unified the opposite behavior of RIS as a scatterer and as a mirror based on the free-space Green function. The derived path loss model is similar to \cite{Ell19}. The results showed that depending on its size and distance, the RIS can be observed as a zero, one, or two dimensional object whose radiated power exposes a dependence with the fourth, third, or second power of the distance, respectively.

In \cite{Ren20_2}, a simple path loss model was proposed for RIS communications based on the general scalar theory of diffraction and the Huygens-Fresnel principle. The RIS was modeled as a sheet of electromagnetic material of negligible thickness. The proposed approach can identify the conditions under which an RIS of finite size can be approximated as an anomalous mirror or as a scatterer.

In \cite{Dan21}, a physics-consistent analytical free-space path loss model for RIS was proposed based on the vector generalization of Green's theorem. The path loss model can be applied to 2D homogenized metasurfaces. The path loss is formulated in terms of a computable integral that depends on the transmission distances, the polarization of the radio waves, the size of the surface, and the desired surface transformation. Closed-form expressions were obtained in far field and near field deployments. 

In \cite{Bou21}, a path loss model was proposed for RIS-assisted THz communication systems, taking into account the THz band and RIS characteristics. Specifically, the proposed path loss model revealed the relationships between the RIS specifications, such as size, number of RIS elements, element's size, reflection coefficient, radiation pattern, phase shift, and the transmission parameters, such as frequency band, Tx-Rx distance, THz specific parameters, and so on. 

In \cite{Wang21_2}, a radar cross section (RCS) based path loss model for RIS communications was proposed. The received power was related to the distances from the Tx/Rx to the RIS, the angles in the Tx-RIS-Rx triangle, and the effective area and the reflection coefficient of each element. In addition, channel measurements were conducted in different scenarios to validate the proposed model in both near field and far field conditions.
    
In \cite{Naj21_1, Naj21_2}, the RIS was partitioned into tiles, where each tile consisted of several unit cells. The far field tile response function $g$ is obtained based on the concept of RCS. Then, the free-space path loss for RIS based link is given as

\begin{equation}
PL_{RIS}=\frac{4\pi \left | g \right |^2 }{\lambda ^{2} } PL_tPL_r
\end{equation}
where $PL_t=(\frac{\lambda }{4\pi d_1} )^2$ and $PL_r=(\frac{\lambda }{4\pi d_2 } )^2$ are the free-space path losses for Tx-RIS and RIS-Rx links, respectively. In addition, the continuous and discrete tile response functions were derived for a given transmission mode. 

In \cite{Cos21}, a similar path loss model was proposed consistent with the path loss model in \cite{Tang21_1} by using the concept of RCS, which is more common for scattering problems. The maximum RCS of the unit cell can be approximated as the gain of the unit cell multiplied by its area.

\begin{figure}[b]
\begin{center}
\vspace{1.5cm}
 \includegraphics[width=3.5in]{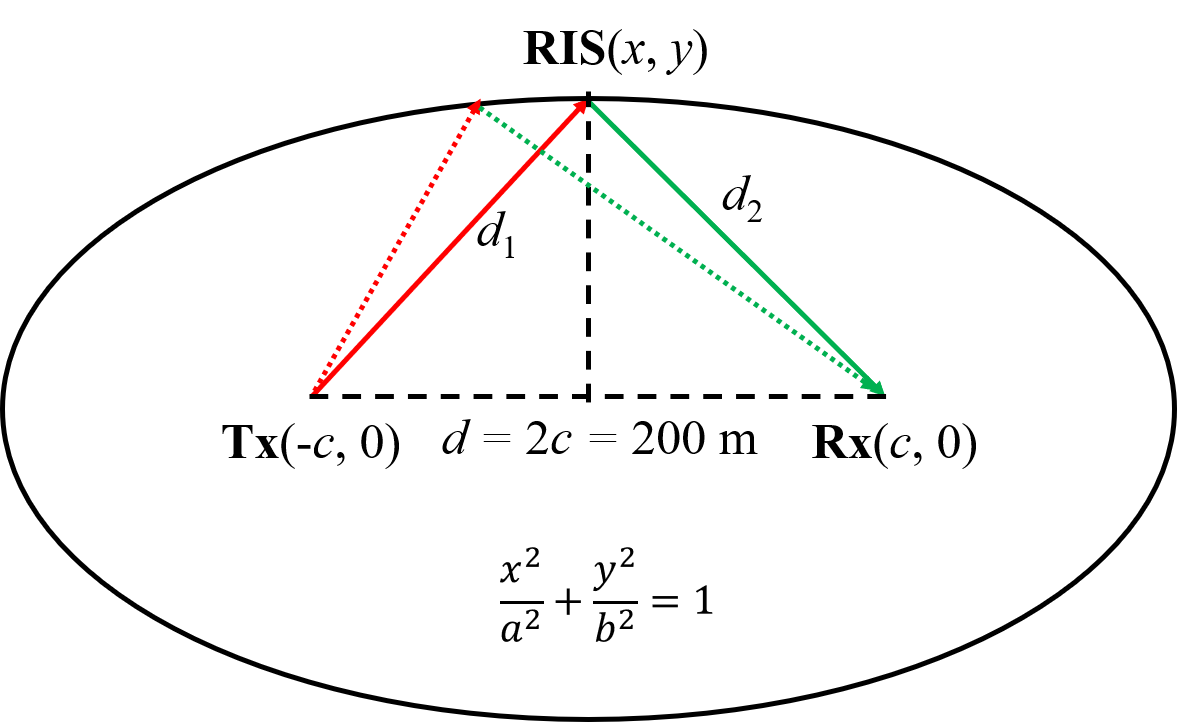}\\
  \caption{A RIS-assisted communication setup with RIS moving on an ellipse.}\label{IRS_pathloss}
\end{center}
\end{figure}

\begin{figure}[!ht]
\begin{center}
\vspace{1.5cm}
 \includegraphics[width=3.5in]{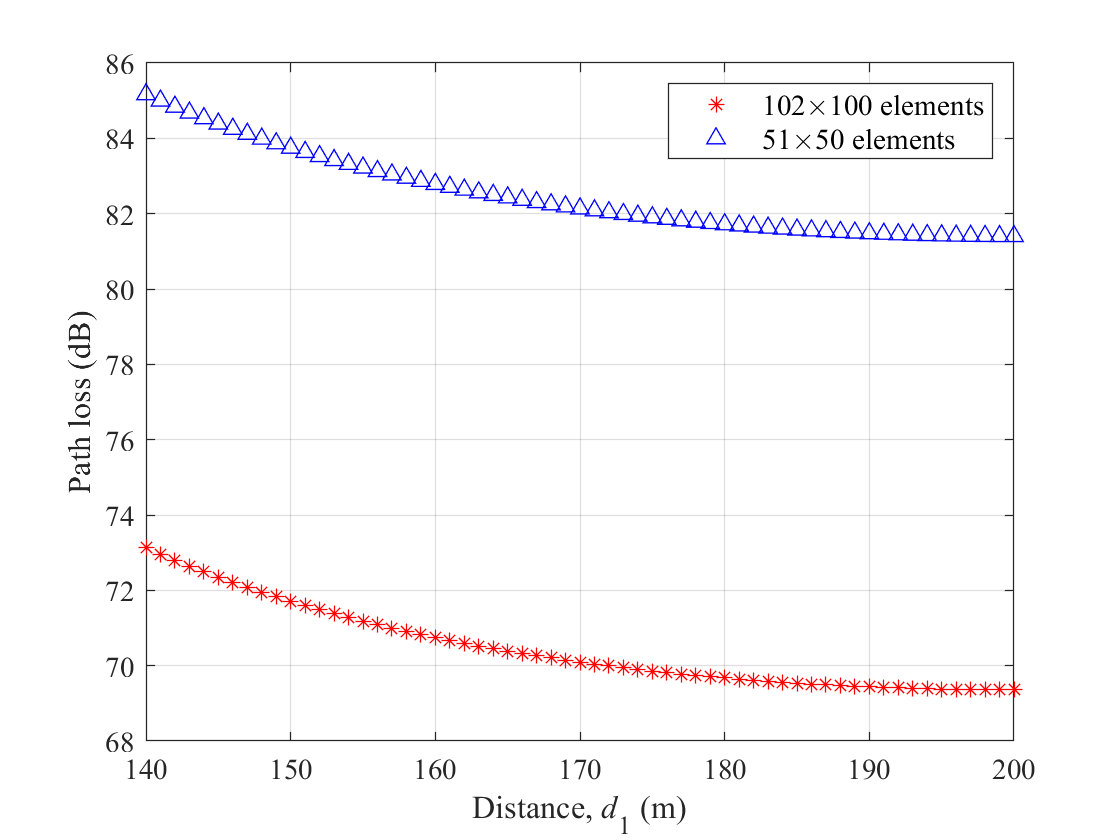}\\
  \caption{Calculated path loss with $d_1$ increasing from 140 m to 200~m.}\label{IRS_pl}
\end{center}
\end{figure}

To illustrate the far field path loss, Fig. \ref{IRS_pathloss} shows an RIS-assisted communication setup. The Tx and Rx are on the focal of an ellipse with 2D coordinates of (-$c$, 0) and ($c$, 0), respectively. The Tx-Rx distance is $d$ = 2$c$ = 200 m. The long semi-axis of the ellipse is $a$ = 200 m. Thus, $b$ = 173 m and $d_1$ + $d_2$ = 400 m. The Rayleigh distance for a 1 m$\times$1.02 m RIS at 10.5 GHz in \cite{Tang21_1} is about 140 m. We simulate the path loss when the RIS moves from $d_1$ = 140~m to $d_1$ = 200 m. The path loss is shown in Fig. \ref{IRS_pl}. As can be seen, the path loss decreases from 73 dB to 69 dB. In case of free space path loss with a distance of $d_1$ + $d_2$ = 400 m, the path loss at 10.5 GHz is 105 dB. If the 21 dB antenna gains at both sides are excluded, the multiplied path losses of $d_1$ and $d_2$ would be larger than the free space path loss. When the size of the RIS reduces from 102$\times$100 to 51$\times$50, the path loss increases about 10 dB, which indicates that a very large size up to ten thousands of RIS elements are needed to combat the path loss.

\begin{figure}[!ht]
\begin{center}
\vspace{1.5cm}
 \includegraphics[width=3.5in]{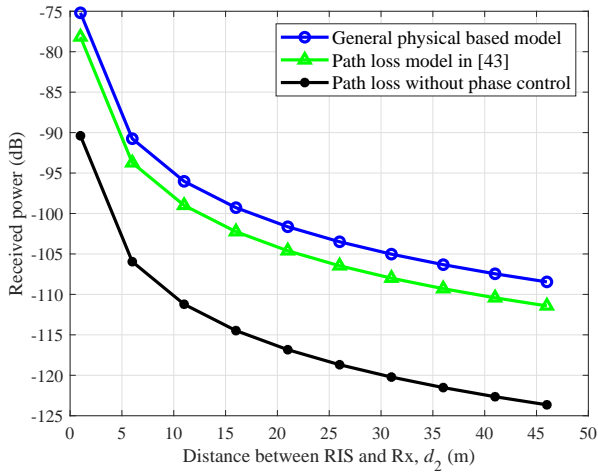}\\
  \caption{Comparison of RIS-assisted path loss with and without phase control.}\label{RIS_pl2}
\end{center}
\end{figure}

In Fig. \ref{RIS_pl2}, we compare the condition of RIS with and without phase control capability. The general physical model in \cite{Naj21_2} and the path loss model in \cite{Tang21_3} are consistent. Note that the received power is relative to a transmitted power of 0 dBm. The simulated frequency band is 5.4 GHz, $M$ = $N$ = 60, $d_x$ = $d_y$ = 0.33$\lambda$, $d_1$ = 20 m. The incident angle is 60$^\circ$ and the reflection angle is 45$^\circ$. For the programmable RIS, each element is set to a different phase as the way in \cite{Tang21_3}. It is clear that the received power is largely improved after the phase control of RIS.

\subsection{Small-scale multipath fading models}

\textcolor{black}{In the literatures, some simple end-to-end fading channel models were used to assess the RIS-assisted wireless communication system performance. The Rician fading model was used for RIS-assisted base station to UAV channel and UAV to ground channel in \cite{Jar21} and for RIS-assisted downlink NOMA cellular system in \cite{Elh22}. In \cite{Ni21}, the performance of RIS-assisted communication system was analyzed under Nakagami-m fading model. In \cite{Tri20}, a generalized fading model was applied to RIS-assisted communication system performance analysis in the presence of phase noise, where the Fox’s H model was utilized as a general fading distribution with Rayleigh and Rician as special cases.}
   
A few RIS-assisted small-scale channel models have been proposed recently, including phase-shift model, transmission-mode model, physics based channel model, beamspace channel model, and geometry based stochastic model (GBSM), which are summarized in Table \ref{tab:model} and thereby overviewed below.

\begin{table*}[tb!]
\caption{A summary of RIS small-scale channel models.}
\label{tab:model}
\centering
  \arraybackslash
\begin{tabular}{|m{1cm}<{\centering}|m{2.5cm}<{\centering}|m{5cm}<{\centering}|m{6.5cm}<{\centering}|}
\hline
\textbf{Ref.}&\textbf{Channel model}&\textbf{Main ideas}&\textbf{Features}\\
\hline
\cite{Naj21_1, Naj21_2}&Phase-shift model&Separate the cascaded channel as three sub-channels &1) Widely use in the literature; 2) RIS properties are not considered.\\
\hline
\cite{Naj21_1, Naj21_2}&Transmission-mode model&1) Partition the large RIS into tiles and calculate tile response function; 2) Low rank channel matrix decomposition; 3) Transmission mode selection&Scalable with RIS elements \\
\hline
\cite{Xu20_1, Xu20_2}&Physics based channel model&Consider the RIS and the scattering environment as a whole&The RIS-aided wireless channel can be approximated by a Rician distribution\\
\hline
\cite{Sun21}&Physics based channel model&Model by the summation of impinging plane waves&Can derive the joint spatial-temporal correlation functions under both isotropic and non-isotropic scattering \\
\hline
\cite{Piz20_1, Piz20_2, Piz21}&Physical channel model&Model the small-scale fading as a zero-mean, spatially-stationary, and correlated Gaussian scalar random field& Satisfy the Helmholtz equation in the frequency domain\\
\hline
\cite{Aja21}&Analytical channel model&Based on the Huygens-Fresnel principle& 1) Apply to RIS-assisted FSO systems; 2) Consider the non-uniform power distribution of Gaussian beams, the impact of RIS size, the positions of the laser source, the lens, and the phase-shift configuration of the RIS\\
\hline
\cite{Ala21}&Beamspace channel model&View RIS as a controlled scattering cluster&Based on the extended SV channel model\\
\hline
\cite{Bas20, Bas21_2, Yil21, Bas21_3}&GBSM&Follow 3GPP channel modeling approach&Can be applied to various indoor/outdoor scenarios and frequency bands\\
\hline
\cite{Dang20}&GBSM&Model the three sub-channels by ellipses&2D, consider the spatial CCFs of sub-channels associated with different RIS elements \\
\hline
\cite{Jiang21}&GBSM&Model the three sub-channels by ellipsoids&3D, Tx and Rx are equipped with multiple antennas \\
\hline
\cite{Sun20, Sun21_2}&GBSM&Model the three sub-channels by GBSMs&Wideband non-stationary properties and practical phase shifts of RIS are studied \\
\hline
\cite{Dov21}&GBSM&Consider the spherical wavefront and THz band path loss&Tx and Rx are equipped with a single antennas \\
\hline
\cite{Zeg20}&Keyhole channel model&Consider the RIS as a
dyad with one degree of freedom&The direct Tx-Rx channel
is ignored. The joint Tx-RIS-Rx channel is modeled as
the sum of the contributions from each RIS element. \\
\hline
\cite{San21}&ML-based channel model&Approximate the fading distributions using EM algorithm&Consider beamforming,
spatial channel correlation, phase-shift errors, arbitrary fading
conditions, and coexistence of direct and RIS channels \\
\hline
\end{tabular}
\end{table*}

\subsubsection{Phase-shift model \& transmission-mode model}
The phase-shift model is widely used in the literature which means the cascaded three sub-channels, i.e., the direct Tx-Rx channel, Tx-RIS channel, and the RIS-Rx channel. However, the phase-shift model has two drawbacks \cite {Naj21_1, Naj21_2}. The first one is that the ranks of Tx-RIS matrix and RIS-Rx matrix are limited by the number of scatterers, which is small compared to the number of RIS elements. The second one is that the physical properties of RIS, the incident and reflected angles, and polarization are not considered. Thus, a transmission-mode model was proposed in \cite {Naj21_2} by three key ideas. Firstly, it partitioned the large RIS into tiles and calculated tile response function. Then, the low rank channel matrices were decomposed. At last, different transmission modes were selected for each tile, where each transmission mode effectively corresponded to a given configuration of the phase shifts that the unit cells of the tile apply to an impinging electromagnetic wave. 

\subsubsection{Physics based channel model}
In \cite{Xu20_1, Xu20_2}, a physics based channel model was proposed for RIS communications. It considers the RIS and the scattering environment as a whole by studying the signal's multipath propagation. The joint channel contains a specular component by the RIS LOS link and a scatter component by the NLOS direct link. Channel statistics were analyzed by considering $M$ RIS MPCs and $N$ MPCs for the scattering environment. The model suggests that the RIS-aided wireless channel can be approximated by a Rician distribution.

In \cite{Sun21}, the Tx-RIS channel was modeled by the summation of impinging plane waves. Then, the joint spatial-temporal correlation functions for RIS under both isotropic and non-isotropic scattering were derived. A closed-form four dimensional (4D) sinc function is derived to describe the joint spatial-temporal correlation in the isotropic scattering environment. For the more practical non-isotropic scattering, a realistic 3D angular distribution extracted from mmWave channel measurements in the urban micro scenario was utilized.

In \cite{Piz20_1, Piz20_2, Piz21}, a spatially-stationary channel model was proposed for holographic MIMO small-scale fading in the far filed case. It modeled the array with a massive number of antennas in a compact space, which is the case of RIS. The small-scale fading was modeled as a zero-mean, spatially-stationary, and correlated Gaussian scalar random field satisfying the Helmholtz equation in the frequency domain, which was equivalent to the scalar wave equation in the time domain. 

In \cite{Aja21}, an analytical channel model was proposed for RIS-assisted free space optical (FSO) systems based on the Huygens-Fresnel principle. Several effects, such as the non-uniform power distribution of Gaussian beams, the impact of RIS size, the positions of the laser source, the lens, and the phase-shift configuration of the RIS were considered. The simulation results validated the accuracy of the proposed analytical model.

\subsubsection{Beamspace channel model}
\begin{figure}[b]
\begin{center}
\vspace{1.5cm}
 \includegraphics[width=3.5in]{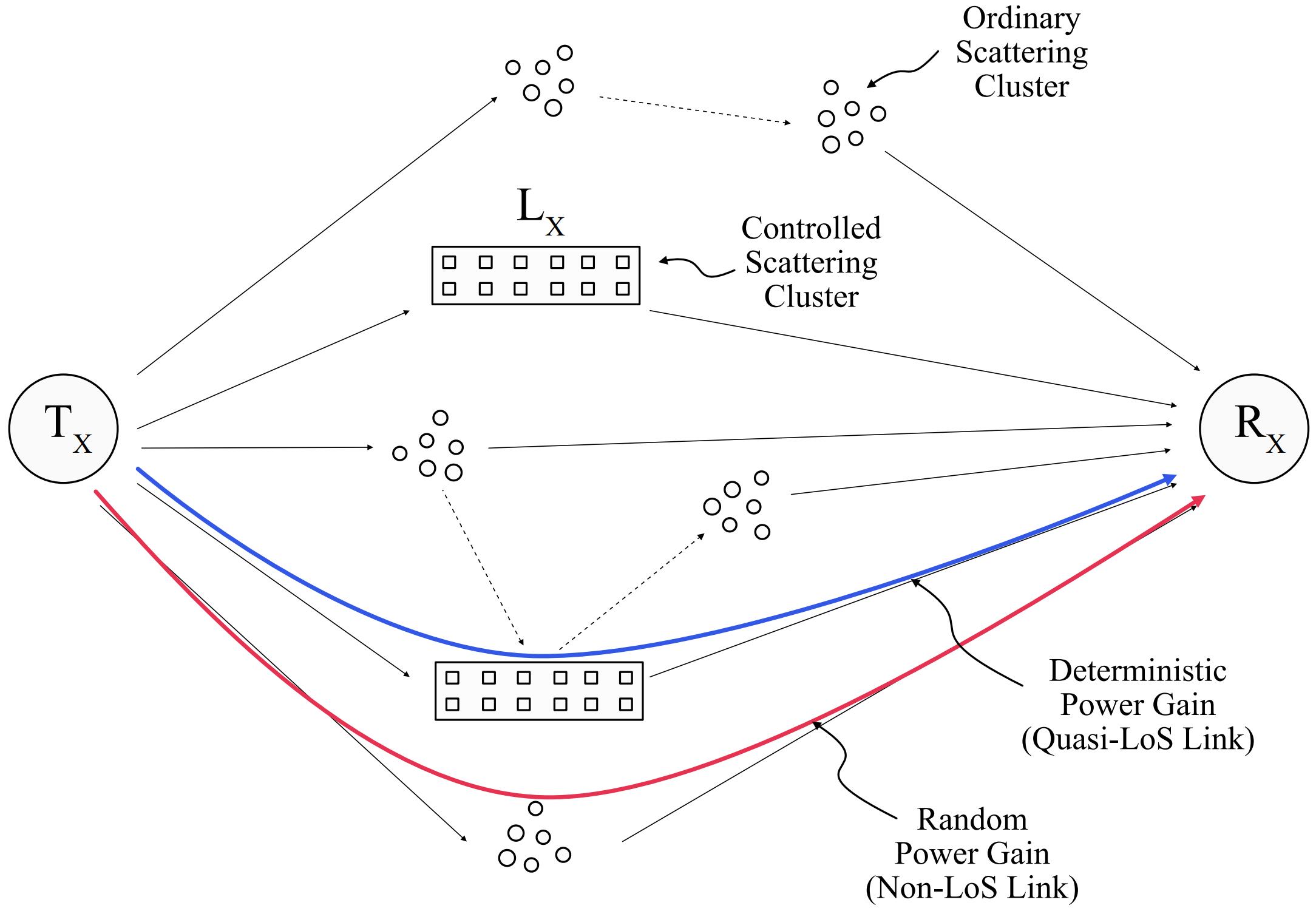}\\
  \caption{A beamspace channel model for RIS-assisted communication scenario \cite{Ala21}.}\label{Beamspace}
\end{center}
\end{figure}

In \cite{Ala21}, a beamspace channel model was proposed for RIS communications, as shown in Fig. \ref{Beamspace}. Different from the phase-shift model, the proposed model viewed RIS as a controlled scattering cluster reflecting its own MPCs. The model is based on the extended Saleh Valenzuela (SV) channel model. An antenna segmentation method was proposed to fit RIS in the developed far field model and showed the characteristics of its MPCs and how they can be controlled. 

\subsubsection{GBSM}
\begin{figure*}[!ht]
\begin{center}
\vspace{1.5cm}
 \includegraphics[width=7in]{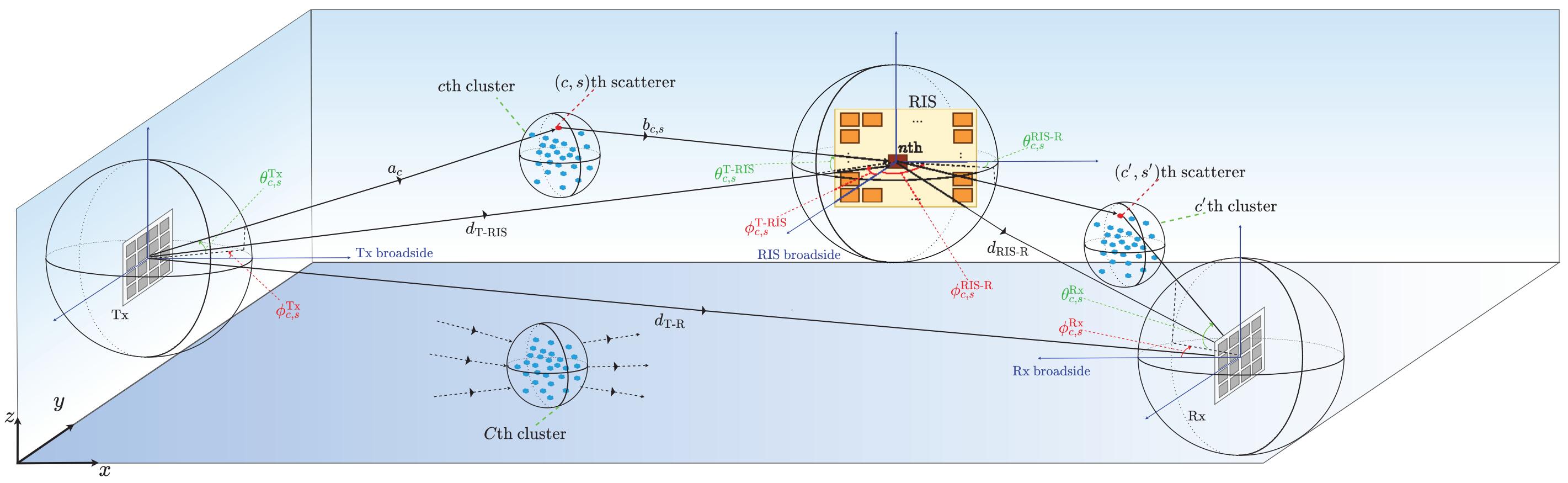}\\
  \caption{A GBSM for RIS-assisted communication scenario \cite{Bas21_3}.}\label{GBSM_Bas}
\end{center}
\end{figure*}

The GBSM approach has been widely used in 5G and 6G channel modeling, such as the proposed channel models in \cite{WuTCOM, BianTWC, WangCOMST, HuangJSAC, HeTC, WangTVT}. In \cite{Bas20, Bas21_2, Yil21, Bas21_3}, an open-source SimRIS channel simulator was presented for RIS-assisted communications, as shown in Fig. \ref{GBSM_Bas}. The proposed channel model was applicable to various indoor and outdoor scenarios and various frequency bands, including mmWave band. The model is similar to 3GPP cluster channel model. It incorporated the LOS probabilities between terminals, array responses of RISs and Tx/Rx units, RIS element gains, realistic path loss and shadowing models, and environmental characteristics in different scenarios and frequency bands. However, it is only applied to far field case. In \cite{Tou21}, the SimRIS channel simulator was adopted to provide performance insights for design of RIS-assisted smart radio environments. The impact of positioning/placement of RISs, the number of reflecting elements and the tilt/rotation of RISs in both single and multi-RIS environments were investigated. Results revealed that these aspects are crucial to achieving capacity and reliability
enhancements in smart radio environments. 

\begin{figure}[!ht]
\begin{center}
\vspace{1.5cm}
 \includegraphics[width=3.5in]{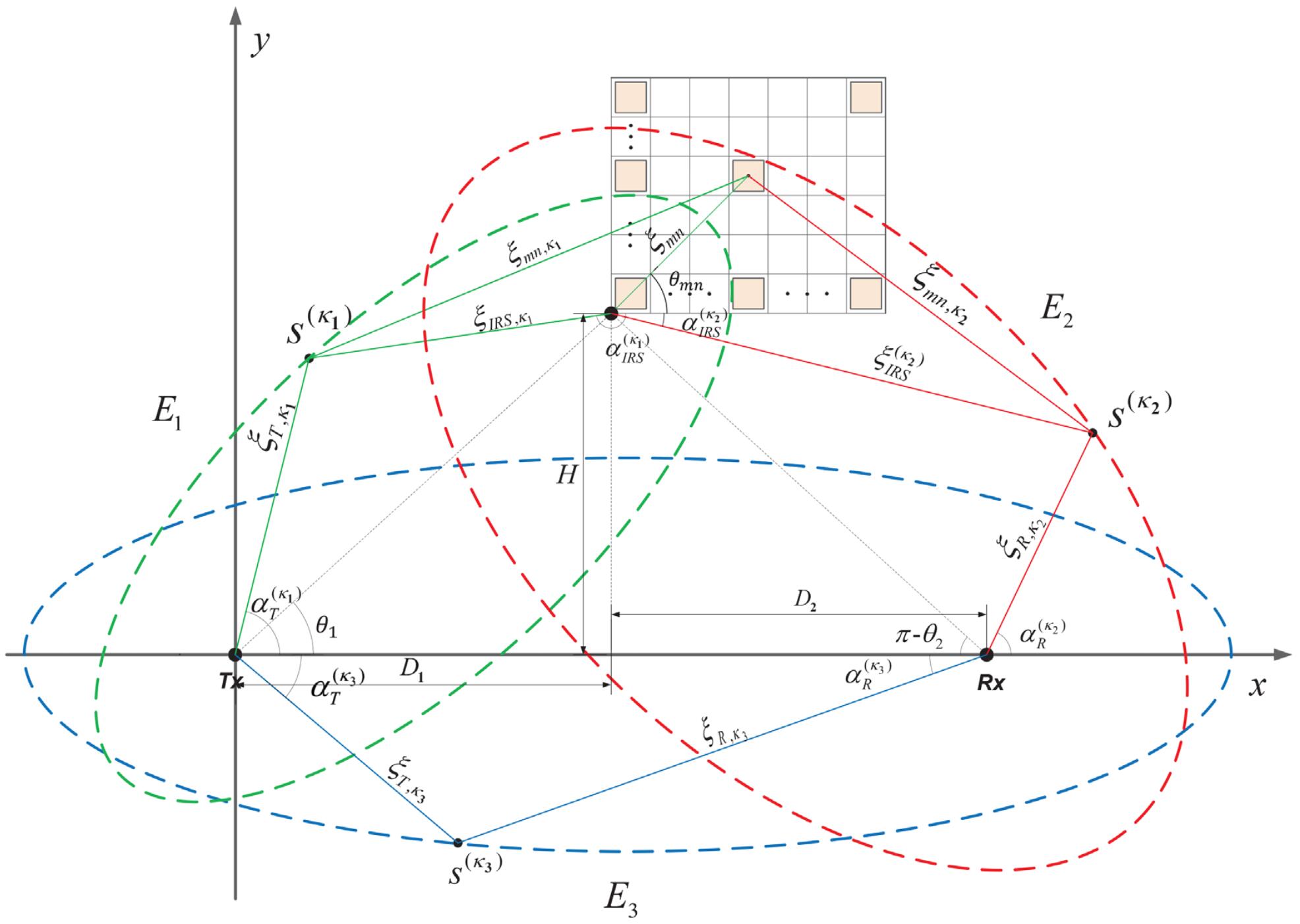}\\
  \caption{A GBSM for RIS-assisted communication scenario \cite{Dang20}.}\label{GBSM_Dang20}
\end{center}
\end{figure}

In \cite{Dang20}, a GBSM was proposed for RIS communications, as shown in Fig. \ref{GBSM_Dang20}, where the three ellipses were used to model Tx-RIS, RIS-Rx, and Tx-Rx channels, respectively. The Tx and Rx were assumed to be equipped with a single antenna, whereas the RIS was a rectangular array. As the proposed model assumed that the Tx, RIS, and Rx were placed in the $x$–$y$ plane of a Cartesian coordinate system, it is a 2D GBSM. The spatial cross-correlation functions (CCFs) of sub-channels associated with different RIS elements were considered.

\begin{figure}[!ht]
\begin{center}
\vspace{1.5cm}
 \includegraphics[width=3.5in]{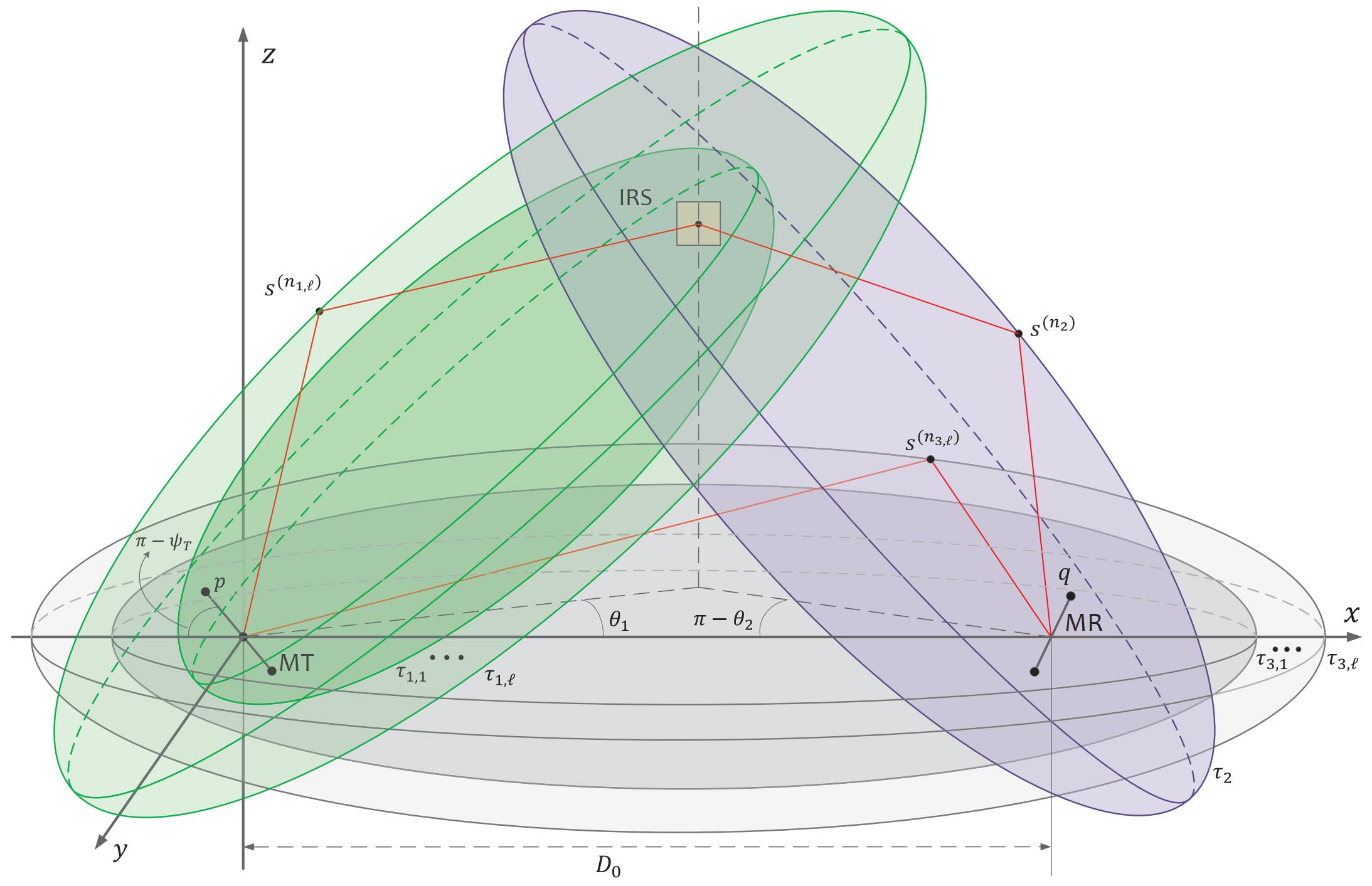}\\
  \caption{A 3D ellipsoid GBSM for RIS-assisted communication scenario \cite{Jiang21}.}\label{GBSM_Jiang21}
\end{center}
\end{figure}

In \cite{Jiang21}, the 2D GBSM proposed in \cite{Dang20} was extended to 3D wideband case by adopting three ellipsoids, as shown in Fig. \ref{GBSM_Jiang21}. The Tx and Rx were equipped with multiple antennas. The non-stationary property was considered. Statistical properties, such as temporal autocorrelation function (ACF), spatial CCF, and frequency correlation function (FCF) were analyzed.

In \cite{Sun20}, we proposed a 3D non-stationary RIS-assisted MIMO GBSM. The twin cluster concept was applied. The channel model supports movements of Tx, Rx, and clusters. The RIS-assisted channel is divided into three sub-channels, i.e., Tx-RIS channel, RIS-Rx, and Tx-Rx channel. The path loss model in \cite{Tang21_2} was incorporated in the channel coefficients. The temporal ACF and spatial CCF were analyzed.

\begin{figure}[!ht]
\begin{center}
\vspace{1.5cm}
 \includegraphics[width=3.5in]{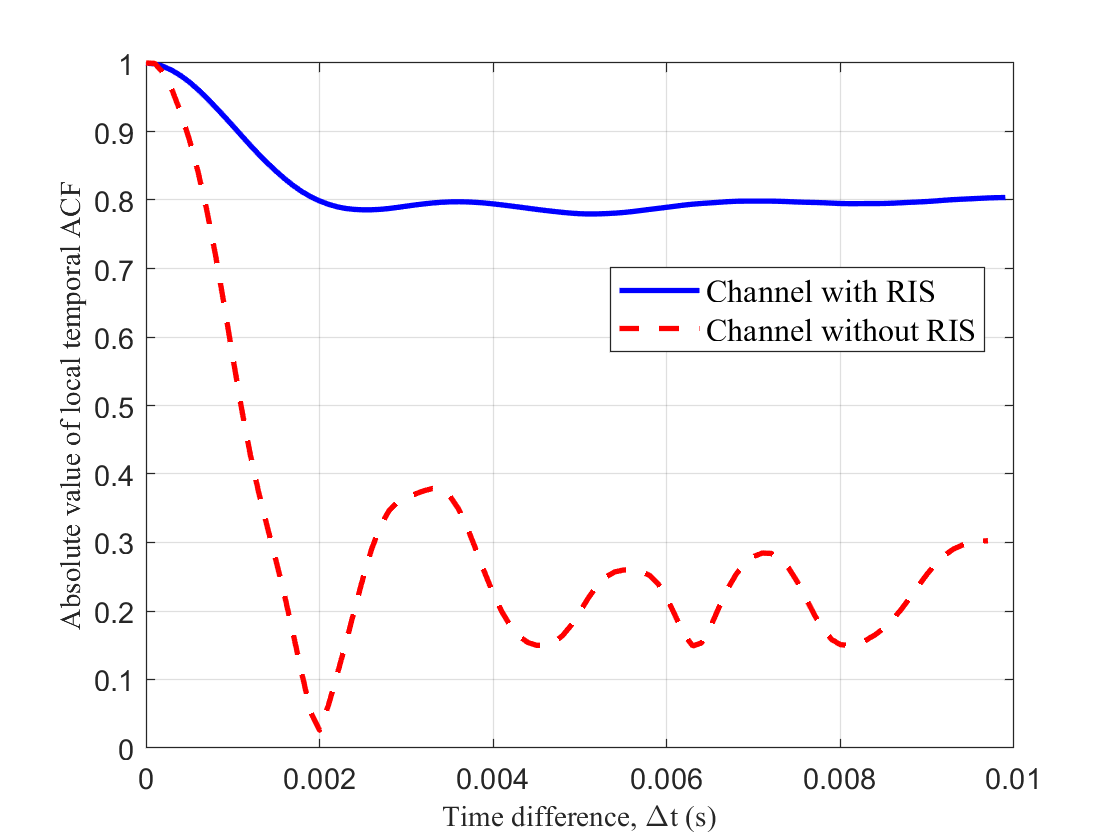}\\
  \caption{Comparison of temporal ACFs with and without RIS.}\label{IRS_ACF}
\end{center}
\end{figure}

We further take the practical discrete RIS phase shifts into account in \cite{Sun21_2}. The cluster evolution in the space domain and practical discrete phase shifts are considered. The statistical properties, including temporal ACF, spatial CCF, local Doppler spread, and RMS delay spread are investigated. We find that RIS plays a role in separating the whole channel. We take the temporal ACF as an example to compare channel statistical properties with and without RIS, as shown in Fig. \ref{IRS_ACF}. It can be seen that the absolute value of temporal ACF employing RIS is much larger than the situation without RIS. In fact, temporal ACF reflects the correlation among different time instances. A larger value indicates that channel is more stable with RIS. The phenomenon proves that RIS can improve the stability of channel. This is because RIS can eliminate the phase differences among LOS components and the strength of the received signal is always strong at different time instances. 

In \cite{Dov21}, a RIS-assisted channel model was proposed for THz band. The Tx and Rx were assumed to be equipped with a single antenna. The spherical wavefront, i.e., the phase difference between adjacent RIS elements was considered. The path loss model for THz bands was employed, which relied on the plate scattering paradigm and molecular absorption. The simulation results showed that a RIS can significantly improve the energy efficiency of MIMO system in near field and beamfocusing.

\textcolor{black}{Other channel models have also been proposed for different scenarios based on GBSM approach, e.g., outdoor \cite{Yang_ICECE}, fixed-to-mobile \cite{Sun_VTC}, double-RIS \cite{Jiang_WCSP}, high altitude platform \cite{Lian_TVT}, UAV \cite{Cao_CL, Jiang_WCL, Xiong_TWC}, and general MIMO \cite{Xiong_TC} scenarios.}

\subsubsection{Keyhole channel model}
In the concept of keyhole MIMO channel, the channel was considered as a dyad with one degree of freedom, where each RIS element was a perfect example of keyhole. In \cite{Zeg20}, the RIS-MIMO channel was modeled as a keyhole MIMO channel. The direct Tx-Rx channel was ignored. The joint Tx-RIS-Rx channel was modeled as the sum of the contributions from each RIS element. Based on that, a channel estimation framework was proposed using single value decomposition (SVD) to separate the cascaded channel links and estimate each link separately. \textcolor{black}{Compared to the phase shift model of RIS, this method considers the amplitude response of RIS element, making the channel model more accurate. With the number of RIS elements increasing, the dimension of Tx-RIS channel coefficient matrix and RIS-Rx channel coefficient matrix will increase linearly, and the complexity of this channel model also increases.}

\subsubsection{ML-based channel model}
ML-based approach has been a new method in channel modeling \cite{TBD20}. \textcolor{black}{Compared to the traditional channel modeling methods, the most important two advantages are that it can predict channel characteristics by training channel measurement data and can explore the complex nonlinear relationships between channel characteristics, frequency band, scenario, and RIS-assisted system configurations.} In \cite{San21}, the unsupervised machine learning (ML) algorithm, i.e., expectation-maximization (EM) algorithm was applied to \textcolor{black}{RIS-assisted} channel modeling. Several aspects were considered, including beamforming, spatial channel correlation, phase-shift errors, arbitrary fading conditions, and coexistence of direct and RIS channels. With approximation of a mixture of two Nakagami-m
distributions, the EM algorithm was used to estimate the mixture weight, mean powers, and fading parameters, which can characterize the RIS's composite channels as correlated/independent over conventional/generalized fading paths in the presence of phase errors.

\section{Future Research Directions}
\subsection{RIS channel measurements and parameters estimation}
Current experiments are mainly conducted to demonstrate RIS-assisted wireless communication systems and transceiver designs. Most of the experiments are conducted in simple environments with a short distance, which is far away from reality. Only path loss channel measurements were conducted in an anechoic chamber, while the multipath fading channel measurements are missing. The physical properties of RIS will have great effects on the incident signals. However, RIS can only tune part of the wireless channel, i.e., the MPCs which interact with RIS. Other MPCs, such as the LOS component and reflected MPCs by other scatterers in the environments will have no change. Although channel characteristics have been theoretically analyzed, to further validate the theoretical analysis, RIS experiments and channel measurements are indispensable.  

In RIS-assisted communication scenario, the MPCs which interact with RIS will be affected by the physical properties of RIS. Other MPCs reflected by other scatterers will be independent with RIS. We designed an algorithm to identify whether a MPC interacts with RIS is carried out by investigating this MPC's responses under the regulation of different RIS transmission modes. Similar to the space-alternating generalized expectation-maximization (SAGE) algorithm, the maximum likelihood estimation and grouped coordinate ascent method are used to estimate MPCs' general parameters, i.e., the amplitude, delay, angle of arrival (AoA), angle of departure (AoD), and Doppler frequency. For MPCs that interact with RIS, an extra and novel maximum likelihood estimation for the incident angle and reflect angle at the RIS is considered in each iteration of the algorithm. This extra estimation is realized based on these MPCs' amplitudes under the regulations of different RIS transmission modes \textcolor{black}{\cite{XuICC}}. 

\begin{figure}[!ht]
\begin{center}
\vspace{1.5cm}
 \includegraphics[width=3.5in]{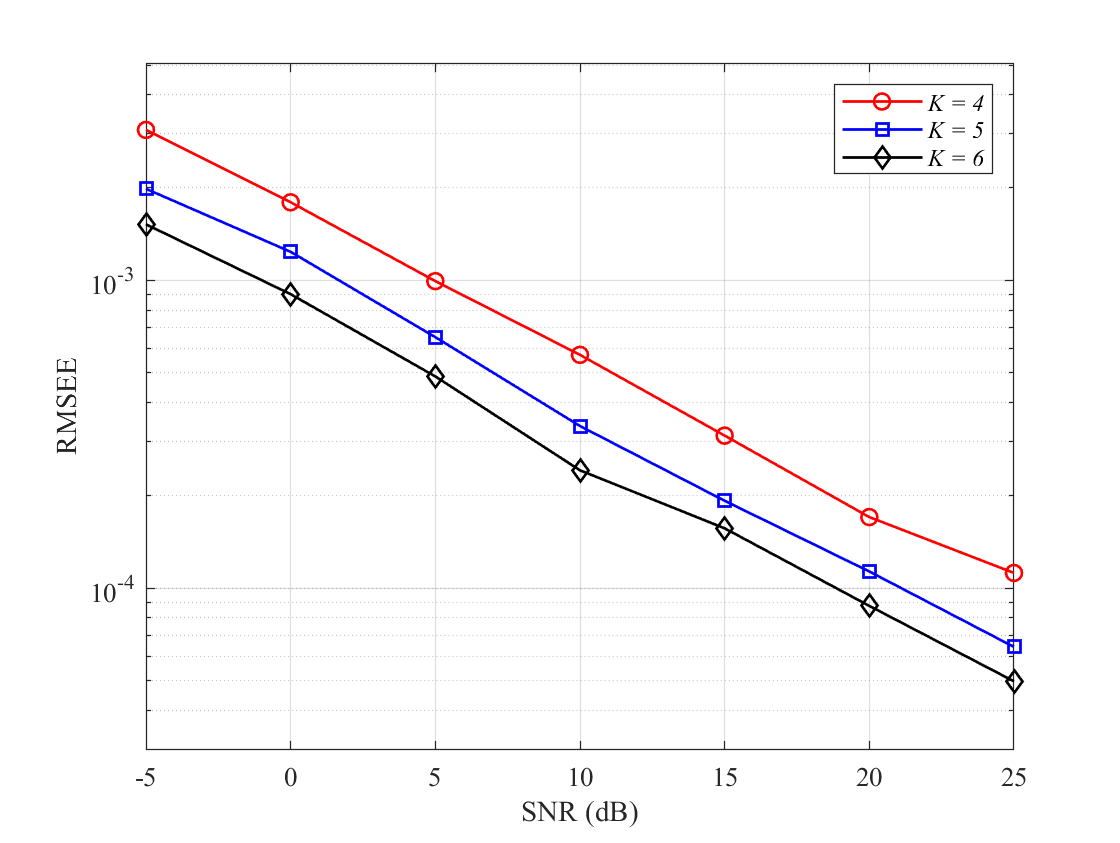}\\
  \caption{RMSEE for different SNRs with different numbers of transmission modes K = 4, 5, and 6.}
  \label{Estimation}
\end{center}
\end{figure}

The root mean square estimation error (RMSEE) for different SNRs and for different number of RIS transmission modes is shown in Fig. \ref{Estimation}. It can be seen that RMSEE reduces with an increase in the number of RIS transmission modes, meaning that more RIS transmission modes lead to higher algorithm's accuracy.

\subsection{A pervasive RIS-assisted channel model}
None of the mentioned RIS-assisted channel models are general enough to take all channel characteristics into account and can be applied to various frequency bands and scenarios. Different channel models will lead to underestimates or overestimates of RIS-assisted communication system performance evaluation. This situation makes it difficult to acquire a standardized RIS-assisted channel model which is pervasive and can be an add-on component. \textcolor{black}{A pervasive RIS-assisted channel model can apply unified channel modeling framework such as the GBSM method, use the unified expression of CIR, and integrate different channel characteristics brought by RIS. By adjusting channel model parameters, the pervasive RIS-assisted channel model is able to adapt to various frequency bands, scenarios, and RIS configurations.} It is also vital for further performance analysis. The new proposed circuit-based channel model considering antenna size and mutual coupling can be a promising method for RIS-assisted channel modeling \cite{GaoWCSP}.

\subsection{AI enabled RIS channel modeling}
Artificial intelligence (AI) has now been widely applied to wireless communications and channel modeling, including MPC clustering, scenario classification, and channel statistical properties prediction \cite{Wang20}. However, AI enabled wireless channel modeling is still not fully explored, especially AI enabled RIS channel modeling. Recent works have shown that generative adversarial network (GAN) is able to learn the end-to-end wireless channels in some simple scenarios \textcolor{black}{\cite{Yang19, Ye20, Sun21_3, Xiao22}. Different from the traditional channel, RIS-assisted channel is a cascade of three sub-channels, i.e., Tx-Rx, Tx-RIS, and RIS-Rx channels. The inherent nonideal channel characteristics depend on the specific analysis of the three sub-channels. However, as RIS is passive, it is not able to obtain the solely responses of Tx-RIS and RIS-Rx channels from traditional channel measurements. Currently, there are no open RIS-assisted channel datasets, which would limit the application of AI enabled RIS channel modeling.}  

\subsection{RIS and electromagnetic information theory}
\textcolor{black}{In principle, the current channel models are for propagation channel, while the radio channel model is lacking. The radio channel takes into account the effects of Tx and Rx antennas. A novel circuit-based radio channel model was proposed in \cite{GaoWCSP}, which can be applied to RIS-assisted radio channel modeling. Due to its unique characteristics, RIS can be a bridge between electromagnetic theory and information theory, building the foundation of electromagnetic information theory. Here, some challenges are pointed out.}  

Firstly, the essence of RIS amplitude and phase control is actually realized by the circuit behind it, so the RF link may need more consideration. The certain circuit parameters may be included when obtaining channel characteristics or calculating the system capacity. Secondly, there is mutual coupling between a large number of RIS sub-wavelength units. Therefore, the entire antenna array can be regarded as continuous when compactly arranged (such as holographic MIMO). This requires starting from Maxwell’ s electromagnetic theory to determine the assumptions and applications of the unit from discrete to continuous. An accurate and easy-to-handle model that can describe the function of the electromagnetic characteristics of the RIS is needed. Thirdly, analysis and optimization of point-to-point RIS wireless networks is an open task, especially for optimizing the phase of the surface and system capacity under different configurations.

\subsection{Applications of RIS channel research to RIS-assisted communications}
The RIS channel research will be essential for RIS-assisted communications, since the performance of RIS-assisted communication systems are largely affected by the underlying channel characteristics, which is summarized in Table \ref{tab:application}.

Take the channel capacity as an example. The channel capacity is a strict upper limit of the rate at which information can be reliably transmitted on the communication channel. RIS can improve the channel capacity by enhancing the received signal power. The ergodic capacity of RIS-assisted single-input and single-output (SISO) system was analyzed in \cite{Bou20}. In \cite{Zhang20} and \cite{Du21}, the channel capacity of RIS-assisted MIMO system in the unicast scenario and multicast scenario were studied, respectively. The alternating optimization algorithm was used to jointly optimize the phase shift matrix of RIS and covariance matrix of the transmitted signal vector. The numerical results illustrated that the channel capacity increased logarithmically with the number of antennas and the square of the number of reflecting elements and it tended to be a constant when both of them went to infinity at a fixed ratio. However, it is inversely proportion to the user number.   Degree of freedom (DoF) offered by RIS will be limited by the number of elements of RIS, since it is impractical to create a RIS containing infinite reflecting elements. To solve this gap, authors in \cite{Su20} proposed irregular RIS, which added a selection matrix in the channel transform function. 

Current analysis of RIS-assisted communications is mainly based on the simplification of RIS and the wireless channel, thus enabling closed-form expressions. More accurate RIS based channel characteristics and channel models can help to improve the design and performance analysis of RIS-assisted wireless communication systems. 

\begin{table}[tb!]
\caption{Applications of RIS channel research to RIS-assisted communications.}
\label{tab:application}
\centering
  \arraybackslash
\begin{tabular}{|m{2.8cm}|m{4.8cm}|}
\hline
\textbf{Applications}&\textbf{Related channel characteristics}\\
\hline
Network optimization and planning& Path loss, shadowing, penetration loss
\\
\hline 
Channel estimation & Delay/angular/Doppler spreads, spatial/temporal/frequency correlation functions\\
\hline 
Signal modulation & Spatial/temporal/frequency correlation functions\\
\hline 
Channel capacity& Path loss, shadowing, spatial/temporal/frequency correlation functions\\
\hline 
Localization/positioning, integrated sensing and communication, communication security& MPC parameters (power/delay/angle), delay/angular/Doppler spread, Ricean K-factor\\
\hline 
\end{tabular}
\end{table}

\subsection{Advanced hardware of RISs}

In most RISs reported above, their main functions are focused on reconfiguring the beam directions and coverages. Recently, some advanced metasurfaces have been developed, such as digital coding and programmable metasurfaces \cite{Cui14_Light, Liu16_Adv, Li17_Nature, Ma17_Adv, Liu17_J, Zhang17_Acs, Shuang20_J, Zhang20_Nature}, which can not only manipulate the beam directions and coverages, but also control the number of beams \cite{Cui14_Light, Liu16_Adv, Li17_Nature}, polarization states \cite{Ma17_Adv}, and waveforms (e.g. cone beams \cite{Liu17_J} and orbital-angular-momentum vortex beams \cite{Zhang17_Acs, Shuang20_J, Zhang20_Nature}) in real time and programmable way. The introduction of time-domain coding increases an additional degree of freedom to control the frequency spectra of electromagnetic waves \cite{Zhao19_National, Dai18_Light}. Hence space-time-coding digital metasurfaces can manipulate both spatial waves and their spectra simultaneously and independently \cite{Zhang18_Nature, Zhang19_Adv}. More importantly, the digital coding metasurfaces make a bridge between the digital world and physical world, and hence they can be used to process the digital information directly, resulting in information metasurfaces \cite{Cui16_Light, Cui17_J, Cui20_Sci}. The information metasurfaces have been applied to build new-architecture wireless communication transmitter systems \cite{Dai19_Adv, Wan19_Light, Cui19_Research, Dai20_TAP, Zhao20_Nature, Zhang21_Nature, Chen21_National}. Based on the information metasurfaces, new electromagnetic information theory have also been presented by considering both electromagnetic waves and bit-stream information simultaneously \cite{Wu20_National, Wu20_Light}. The digital coding and programmable metasurfaces, space-time-coding digital metasurfaces, and information metasurfaces can be regarded as the advanced RISs. In the future, investigating the channel characteristics and channel models of such advanced RISs may produce revolutionary developments of the wireless communications.

\section{Conclusions}
In this paper, we have provided a comprehensive survey on state-of-the-art RIS experiments and channel measurements, channel characteristics, and large-scale path loss models and small-scale multipath fading channel models. Although various RIS prototypes have been made, the experiments and especially channel measurements are still in the infancy. RIS based communications bring several merits to wireless channel characteristics, including simultaneous reflection and transmission, Doppler effect and multipath fading mitigation, channel reciprocity, channel hardening, rank improvement, etc. The large-scale path loss models and small-scale multipath fading models have been compared. In addition, we have pointed out future directions which need further investigation, i.e., RIS prototypes, experiments, and channel measurements, pervasive RIS channel models, AI enabled RIS channel modeling, RIS and electromagnetic information theory, applications of RIS channel research to RIS-assisted communications, and advanced hardware of RISs.

\begin{IEEEbiography}[{\includegraphics[width=1in,height=1.25in,clip,keepaspectratio]{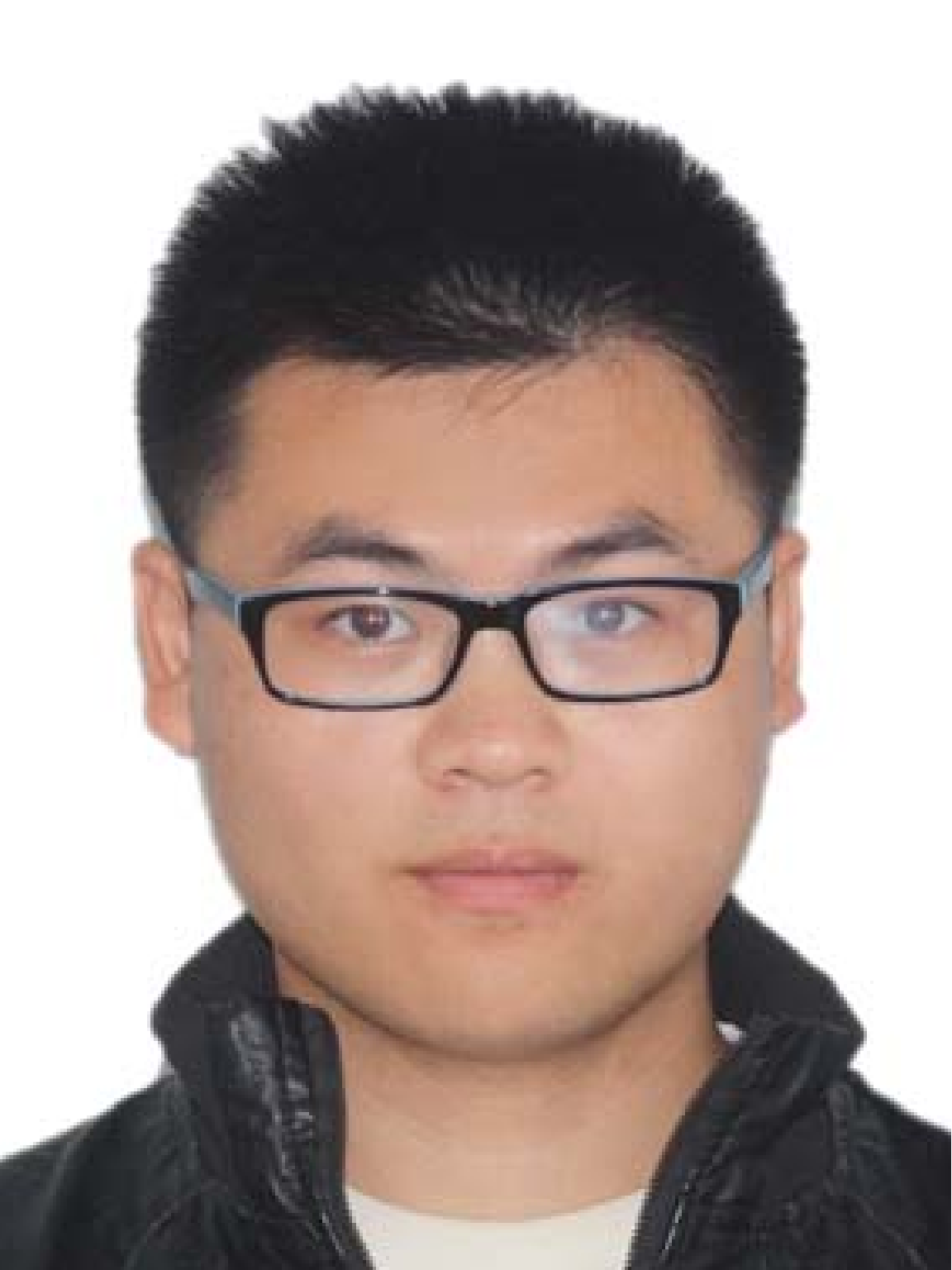}}]{Jie Huang} (M'20) received the B.E. degree in Information Engineering from Xidian University, China, in 2013, and the Ph.D. degree in Communication and Information Systems from Shandong University, China, in 2018. From October 2018 to October
2020, he was a Postdoctoral Research Associate in the National Mobile Communications Research Laboratory, Southeast University, China, supported by the National Postdoctoral Program for Innovative Talents. From January 2019 to February 2020, he was a Postdoctoral Research Associate in Durham University, UK. He is currently an Associate Professor in the National Mobile Communications Research Laboratory, Southeast University, China and also a researcher in Purple Mountain Laboratories, China. His research interests include millimeter wave, THz, massive MIMO, reconfigurable intelligent surface channel measurements and modeling, wireless big data, and 6G wireless communications. He received Best Paper Awards from WPMC 2016, WCSP 2020, and WCSP 2021.
\end{IEEEbiography}

\begin{IEEEbiography}[{\includegraphics[width=1in,height=1.25in,clip,keepaspectratio]{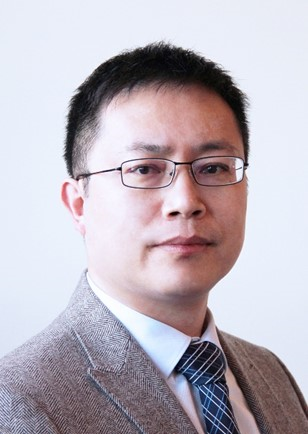}}]{Cheng-Xiang Wang} (S'01-M'05-SM'08-F'17) received the BSc and MEng degrees in	Communication and Information Systems from	Shandong University, China, in 1997 and 2000, respectively, and the PhD degree in Wireless Communications from Aalborg University, Denmark, in 2004.

He was a Research Assistant with the Hamburg University of Technology, Hamburg, Germany, from 2000 to 2001, a Visiting Researcher with Siemens AG Mobile Phones, Munich, Germany, in 2004, and a Research Fellow with the University of Agder, Grimstad, Norway, from 2001 to 2005. He has been with Heriot-Watt University, Edinburgh, U.K., since 2005, where he was promoted to a Professor in 2011. In 2018, he joined the National Mobile Communications Research Laboratory, Southeast University, China, as a Professor. He is also a part-time professor with the Purple Mountain Laboratories, Nanjing, China. He has authored four books, three book chapters, and more than 430 papers in refereed journals and conference proceedings, including 25 Highly Cited Papers. He has also delivered 22 Invited Keynote Speeches/Talks and 9 Tutorials in international conferences. His current research interests include wireless channel measurements and modeling, 6G wireless communication networks, and applying artificial intelligence to wireless networks.
	
Prof. Wang is a member of the Academia Europaea, a fellow of the IET, an IEEE Communications Society Distinguished Lecturer in 2019 and 2020, and a Highly-Cited Researcher recognized by Clarivate Analytics in 2017-2020. He is currently an Executive Editorial Committee member for the IEEE TRANSACTIONS ON WIRELESS COMMUNICATIONS. He has served as an Editor for nine international journals, including the IEEE TRANSACTIONS ON WIRELESS	COMMUNICATIONS from 2007 to 2009, the IEEE TRANSACTIONS ON VEHICULAR TECHNOLOGY from 2011 to 2017, and the IEEE TRANSACTIONS ON COMMUNICATIONS from 2015 to 2017. He was a Guest Editor for the IEEE JOURNAL ON SELECTED AREAS IN COMMUNICATIONS, Special Issue on Vehicular Communications and Networks (Lead Guest Editor), Special Issue on Spectrum and Energy Efficient Design of Wireless Communication Networks, and Special Issue on Airborne Communication Networks. He was also a Guest Editor for the IEEE TRANSACTIONS ON BIG DATA, Special Issue on Wireless Big Data, and is a Guest Editor for the IEEE TRANSACTIONS ON COGNITIVE COMMUNICATIONS AND NETWORKING, Special Issue on Intelligent Resource Management for 5G and Beyond. He has served as a TPC Member, a TPC Chair, and a General Chair for more than 80 international conferences. He received 12 Best Paper Awards from IEEE GLOBECOM 2010, IEEE ICCT 2011, ITST 2012, IEEE VTC 2013-Spring, IWCMC 2015, IWCMC 2016, IEEE/CIC ICCC 2016, WPMC 2016, WOCC 2019, IWCMC 2020, and WCSP 2020.
\end{IEEEbiography}

\begin{IEEEbiography}
[{\includegraphics[width=1in,height=1.25in,clip,keepaspectratio]{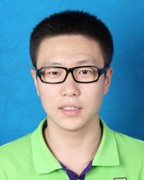}}]{Yingzhuo Sun} received the B.E. degree in Information Engineering from Southeast University, China, in 2020. He is currently pursuing the M.E. degree with the National Mobile Communications Research Laboratory, Southeast University, China. He received the Best Paper Award from WCSP 2020. His research interests include intelligent reflecting surface channel measurements and modeling.
\end{IEEEbiography}

\begin{IEEEbiography}
[{\includegraphics[width=1in,height=1.25in,clip,keepaspectratio]{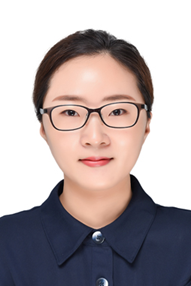}}]{Rui Feng} received the M.Eng. degree in Signal and Information Processing from Yantai University, China, in 2014, and the Ph.D. degree in Information and Communication Engineering from Shandong University, China, in 2018. From July 2018 to Sept. 2020, she was a Lecturer at Ludong University, China. Since Oct. 2020, she has been a Postdoctoral Research Associate in Purple Mountain Laboratories, China. She received the Best Paper Award from WPMC 2016. Her research interests include (ultra-)massive MIMO channel modeling theory and beam domain channel modeling.
\end{IEEEbiography}

\begin{IEEEbiography}
[{\includegraphics[width=1in,height=1.25in,clip,keepaspectratio]{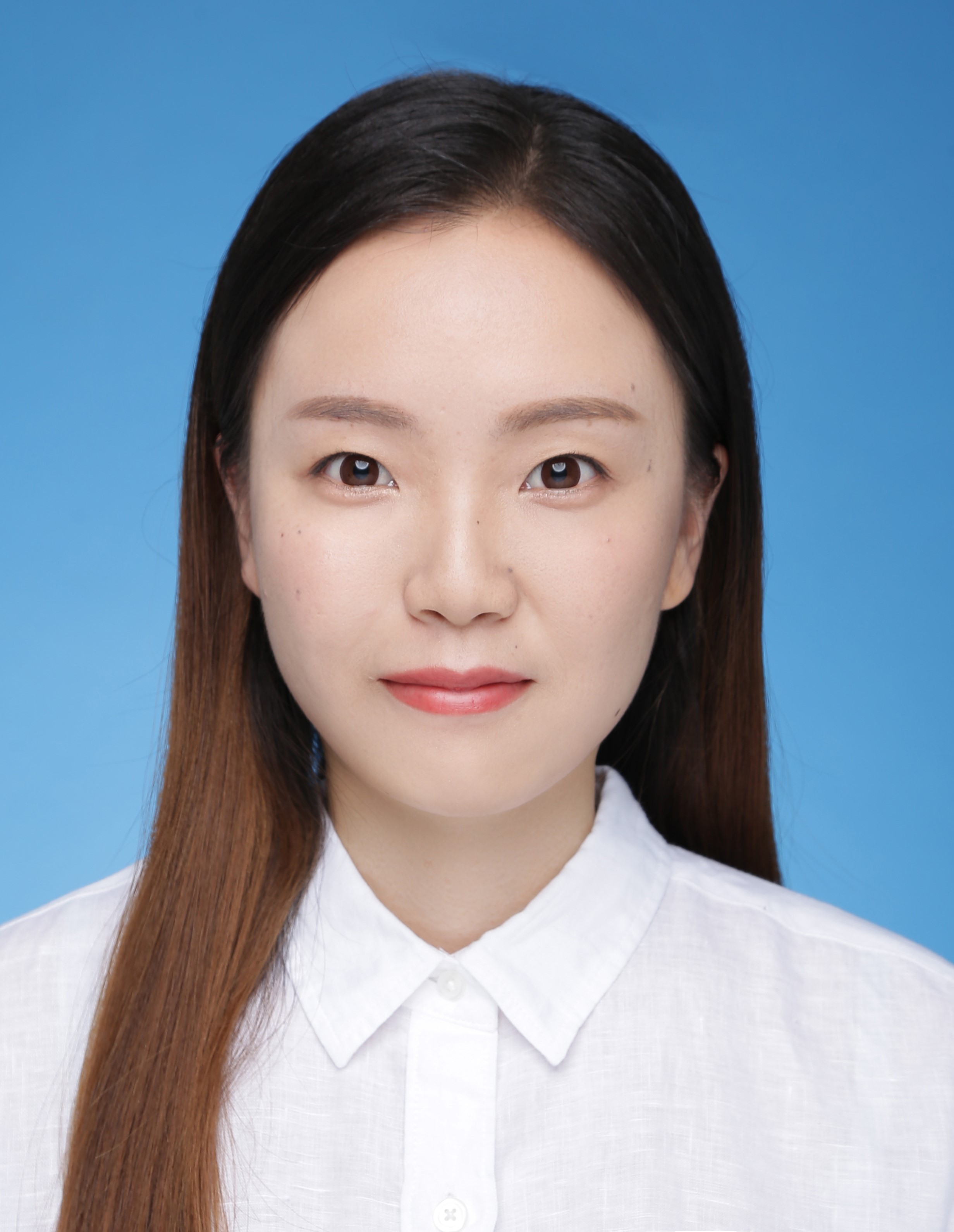}}]{Jialing Huang} received the B.E. degree in Communication Engineering from Soochow University, China, in 2019. She is currently pursuing the M.E. degree with the National Mobile Communications Research Laboratory, Southeast University, China. Her research interests include millimeter wave, THz, and ray tracing channel modeling.
\end{IEEEbiography}

\begin{IEEEbiography}
[{\includegraphics[width=1in,height=1.25in,clip,keepaspectratio]{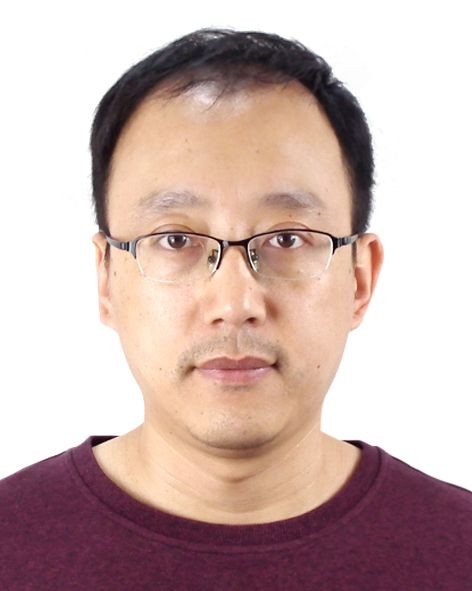}}]{Bolun Guo} received the B.E. degree in Electronic Information Engineering and M.S. degree in Signal and Information Processing from Dalian University of Technologies, China, in 2005 and 2008, respectively. He joined Huawei Technologies Co., Ltd. in 2008. His responsibilities have included ITU and 3GPP standard radio channel model research, 4G/5G system level simulation and evaluation, and spectrum sharing studies.

He has participated in ITU-R WP 3J/3K/3M and SG3 meetings regularly since 2014. He has made considerable contributions to the work of several ITU-R Recommendations. He also participates regularly in ITU-R WP 5D meetings and contributes actively to spectrum sharing studies. In 3GPP, he has contributed to RAN1 channel model standards in 3GPP TR38901, relating to V2V, NTN, IIOT, etc.

His current interests include research into 3GPP/ITU radio propagation characteristics at frequencies that may be considered for systems beyond 5G/6G, fixed links and satellite communication systems, the use of reconfigurable intelligent surface (RIS), and the modelling of radio propagation in spectrum sharing studies. 

\end{IEEEbiography}

\begin{IEEEbiography}
[{\includegraphics[width=1in,height=1.25in,clip,keepaspectratio]{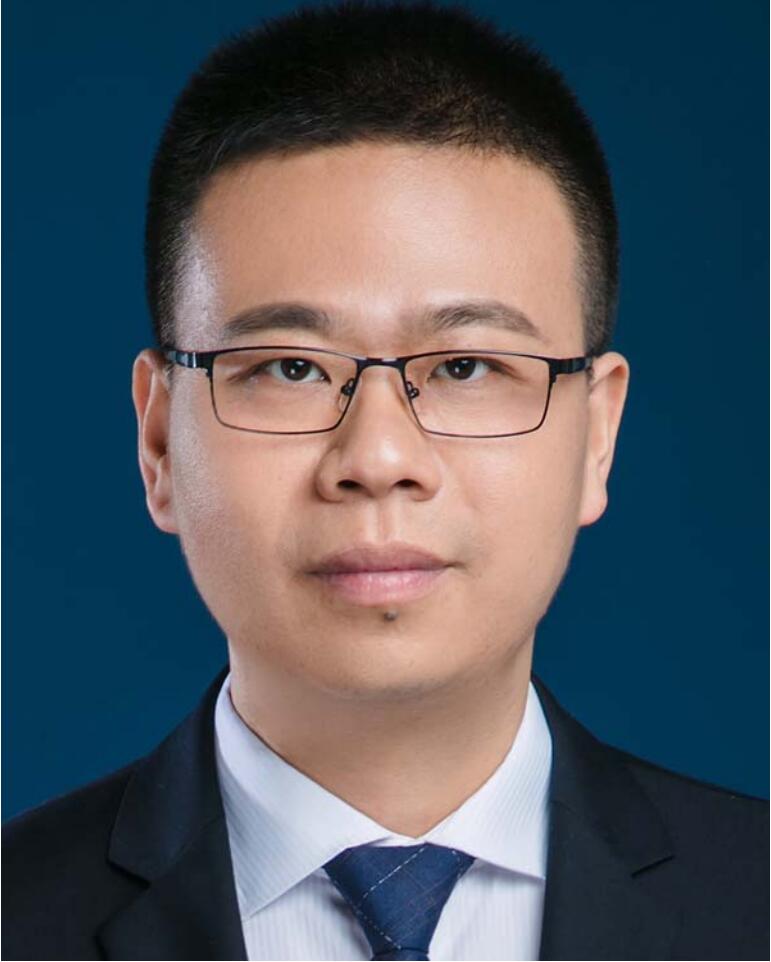}}]{Zhimeng Zhong} received the B.E., M.S., and Ph.D. degrees from Xi'an Jiaotong University, China, in 2002, 2005, and 2008, respectively, all in Electronic Engineering. He joined Huawei Technologies CO., LTD from 2009, and about 11 years experiences on wireless communication system research and development. Dr. Zhong is a senior member of the IEEE. In these years, he mainly focused on wireless channel measurement, channel modeling, and MIMO algorithm \& system design. He is the team leader of wireless channel research in Huawei. Especially, he was in charge of the 3GPP 5G channel model standardization work in Huawei, and gave contributions on 3GPP mmWave channel model, V2V channel model, and IIOT channel model standardization. Recently, he also focused on massive MIMO algorithm design based on some special channel characteristics and machine learning technologies.
\end{IEEEbiography}

\begin{IEEEbiography}
[{\includegraphics[width=1in,height=1.25in,clip,keepaspectratio]{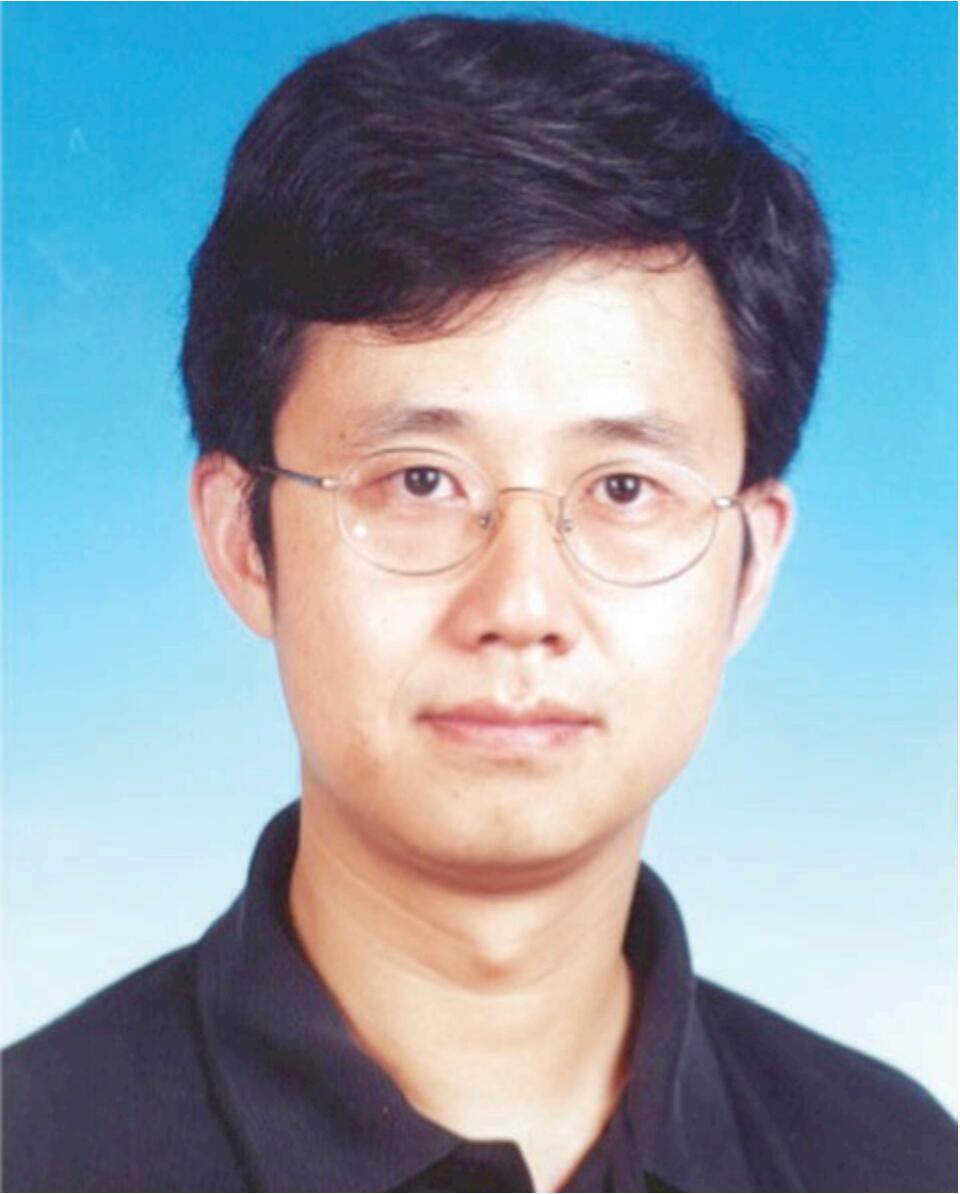}}]{Tie Jun Cui} received the B.Sc., M.Sc., and Ph.D. degrees in electrical engineering from Xidian University, Xi'an, China, in 1987, 1990, and 1993, respectively. In March 1993, he joined the Department of Electromagnetic Engineering, Xidian University, and was promoted to an Associate Professor in November 1993. From 1995 to 1997 he was a Research Fellow with the Institut fur Hochst-frequenztechnik und Elektronik (IHE) at University of Karlsruhe, Germany. In July 1997, he joined the Center for Computational Electromagnetics, Department of Electrical and Computer Engineering, University of Illinois at Urbana-Champaign, first as a Postdoctoral Research Associate and then a Research Scientist. In September 2001, he was a Cheung-Kong Professor with the Department of Radio Engineering, Southeast University, Nanjing, China. Currently he is the Chief Professor of Southeast University. He is also Founding Director of the Institute of Electromagnetic Space, Southeast University.

Dr. Cui's research interests include metamaterials and computational electromagnetics. He proposed the concepts of digital coding and programmable metamaterials, and realized their first prototypes, based on which he founded the new direction of information metamaterials, bridging the physical world and digital world. Dr. Cui is the first author of books Metamaterials – Theory, Design, and Applications (Springer, Nov. 2009); Metamaterials: Beyond Crystals, Noncrystals, and Quasicrystals (CRC Press, Mar. 2016); and Information Metamaterials (Cambridge University Press, 2021). He has published over 500 peer-review journal papers, which have been cited by more than 42000 times (H-Factor 103; Google Scholar), and licensed over 160 patents. Dr. Cui was awarded a Research Fellowship from the Alexander von Humboldt Foundation, Bonn, Germany, in 1995, received a Young Scientist Award from the International Union of Radio Science in 1999, was awarded a Cheung Kong Professor by the Ministry of Education, China, in 2001, and received the National Science Foundation of China for Distinguished Young Scholars in 2002. Dr. Cui received the Natural Science Award (first class) from the Ministry of Education, China, in 2011, and the National Natural Science Awards of China (second class, twice), in 2014 and 2018, respectively. His researches have been selected as one of the most exciting peer-reviewed optics research ``Optics in 2016'' by Optics and Photonics News Magazine, 10 Breakthroughs of China Science in 2010, and many Research Highlights in a series of journals. His work has been widely reported by Nature News, MIT Technology Review, Scientific American, Discover, New Scientists, etc.

Dr. Cui is the Academician of Chinese Academy of Science, and IEEE Fellow. He served as Associate Editor of IEEE Transactions on Geoscience and Remote Sensing, and Guest Editors of Science China - Information Sciences, Science Bulletin, IEEE JETCAS, and Research. Currently he is the Chief Editor of Metamaterial Short Books in Cambridge University Press, the Associate Editor of Research, and the Editorial Board Members of National Science Review, eLight, PhotoniX, Advanced Optical Materials, Small Structure, and Advanced Photonics Research. He presented more than 70 Keynote and Plenary Talks in the Academic Conferences, Symposiums, or Workshops. 

\end{IEEEbiography}

\end{document}